\documentclass[12pt]{elsarticle}

\biboptions{numbers,sort&compress}

\usepackage{amsmath,amssymb}
\usepackage{amsthm}

\usepackage[utf8]{inputenc}
\usepackage[english]{babel}

\usepackage{graphicx}
\usepackage{xcolor}
\usepackage{float}
\usepackage{enumitem}

\usepackage{doi}
\usepackage{hyperref}

\newtheorem{remark}{Remark}

\allowdisplaybreaks

\def\dtp#1{\mathop {#1}\limits_{+\tau}}
\def\dtm#1{\mathop {#1}\limits_{-\tau}}
\def\dsp#1{\mathop {#1}\limits_{+s}}
\def\dsm#1{\mathop {#1}\limits_{-s}}

\newcommand{\ddt}{\partial \over \partial t}

\newcommand{\ddz}{\partial \over \partial z}
\newcommand{\ddr}{\partial \over \partial r}
\newcommand{\ddtheta}{\partial \over \partial \theta}
\newcommand{\dds}{\partial \over \partial s}
\newcommand{\ddu}{\partial \over \partial u}
\newcommand{\ddv}{\partial \over \partial v}
\newcommand{\ddw}{\partial \over \partial w}
\newcommand{\ddrho}{\partial \over \partial \rho }
\newcommand{\ddp}{\partial \over \partial p}

\newcommand{\ddEz}{\partial \over \partial E^z }

\newcommand{\ddHz}{\partial \over \partial H^z }
\newcommand{\ddEtheta}{\partial \over \partial E^{\theta}  }
\newcommand{\ddHtheta}{\partial \over \partial H^{\theta} }

\newcommand{\twopointS}{\widetilde{S}}

\journal{Commun. Nonlinear Sci. Numer. Simul.}

\begin{document}

\begin{frontmatter}

\title{Invariant Finite-Difference Schemes for Cylindrical One-Dimensional~MHD Flows with Conservation Laws Preservation}

\author[mysecondaryaddress]{E.~I. Kaptsov}

\ead{evgkaptsov@math.sut.ac.th}

\author[mymainaddress]{V.~A. Dorodnitsyn}

\ead{Dorodnitsyn@keldysh.ru,dorod2007@gmail.com}

\author[mysecondaryaddress]{S.~V. Meleshko\corref{mycorrespondingauthor}}

\cortext[mycorrespondingauthor]{Corresponding author}

\ead{sergey@math.sut.ac.th}

 \address[mysecondaryaddress]{School of Mathematics, Institute of Science, \\
 Suranaree University of Technology, 30000, Thailand}

 \address[mymainaddress]{Keldysh Institute of Applied Mathematics,\\
 Russian Academy of Science, Miusskaya Pl. 4, Moscow, 125047, Russia}

\date{\today}

\begin{abstract}
On the basis of the recent group classification of the one-dimensional magnetohydrodynamics~(MHD) equations in cylindrical geometry, the construction of symmetry-preserving finite-difference schemes with conservation laws is carried out. New schemes are constructed starting from the classical completely conservative Samarsky--Popov schemes.
In the case of finite conductivity, schemes are derived that admit all the symmetries and possess all the conservation laws of the original differential model, including previously unknown conservation laws. In the case of a frozen-in  magnetic field (when the conductivity is infinite), various schemes are constructed that possess conservation laws, including those preserving entropy along trajectories of motion.
The peculiarities of constructing schemes with an extended set of conservation laws
for specific forms of entropy and magnetic fluxes are discussed.
\end{abstract}

\begin{keyword}
Magnetohydrodynamics
\sep Lagrangian coordinates
\sep conservation laws
\sep Lie point symmetries
\sep numerical scheme
\end{keyword}
\end{frontmatter}

\section{Introduction}

Magnetohydrodynamics (MHD) equations describe the interaction of electrically conductive liquids or gases (electrolytes, liquid metals, plasma) with an electromagnetic field. Important special cases of such interactions are described by one-dimensional equations in plane and cylindrical geometry.
While the authors' papers~\cite{bk:DorKozMelKap_CylFlows_2022} and~\cite{bk:DorKapMDPI2022} are devoted to the case of plane geometry, the present publication is focused on the case of cylindrical geometry.
Flows in magnetohydrodynamic channels, the electrodynamic acceleration of a plasma, magnetic flux ropes behavior,
dynamics of galactic accretion disks and accretion disks around black holes in astrophysics,
and many other phenomena~\cite{Popov1971, Dorodnitsyn1973, Tsui2005,lock_mestel_2008,Suzuki2019}
are studied using the~MHD equations in cylindrical geometry.
Another related phenomenon considered in cylindrical geometry is cylindrical shock waves~\cite{Arora2014,Chauhan2020,Nath2021b,Singh2021}~(see also the references listed in~\cite{bk:DorKozMelKap_CylFlows_2022}).

As the~MHD equations are nonlinear, their integration and finding exact solutions meet difficulties.
Thus, various problems that involve electrically conductive fluids are often studied through numerical simulation,
although the known exact solutions such as~\cite{PandeyManojRadhaSharma2008,Nath2018} are useful both for qualitative analysis of the equations and for testing numerical methods.
A set of methods for numerical modeling of the discussed phenomena has evolved into a separate branch of~MHD
called~Computational magnetohydrodynamics~(CMHD).
Among the large number of the~CMHD approaches~\cite{Toro,bk:SamarskyPopov[1970],MACCORMACK201172,
bk:FalleKomissarovJoarderMHD,bk:POWELL1999284,
bk:YAKOVLEV201380,bk:YANG2017561,bk:HIRABAYASHI2016851,
bk:DongsuMHD}, here we restrict ourselves to considering only finite-difference methods.
In the recent paper~\cite{bk:DorKapMDPI2022}, the authors studied finite-difference schemes for the~MHD equations in plane geometry for finite and infinite conductivity. Further we follow a similar strategy: to take the completely conservative~Samarskiy--Popov schemes~\cite{bk:SamarskyPopov[1970],bk:SamarskyPopov_book[1992]} as a basis, and to extend or modify them so that they possess as much conservation laws as possible. The presence of conservation laws~(energy, momentum, angular momentum, entropy, etc.) is an important property of the equations describing physical phenomena, and is closely related to the presence of symmetries admitted by the equations~\cite{bk:Noether1918,bk:Ibragimov1985}.
Conversely, knowledge of the symmetries of equations allows one to find conservation laws, as well as to derive exact solutions, to reduce partial differential equations to ordinary differential equations~\cite{bk:Ovsyannikov[1962],bk:Ibragimov1985,bk:Olver}.
These statements are also true for finite-difference schemes.

Since the late eighties~\cite{Dor_1,Maeda1,Maeda2}, invariant finite-difference schemes, i.e., schemes that admit symmetries, are intensively studied. A number of methods and techniques for constructing invariant finite-difference schemes, including those with conservation laws, have been developed~\cite{bk:Dorodnitsyn[2011],[Pavel2],
bk:DorodKozlovWint[2004],bk:Dorod_Hamilt[2011],
bk:DorodKozlovWintKaptsov[2015],bk:ChevDorKap2020,bk:LeviWinternitzYamilovBookInPrep2022}.
In practice, for construction of invariant schemes for various partial differential equations of continuum mechanics, the finite-difference analogue of the direct method~\cite{bk:ChevDorKap2020,DorKapSWJMP2021} in combination with the method of difference invariants~\cite{bk:Dorodnitsyn[2011]} is the most commonly used by the authors.
Recently, using these methods, the authors have constructed invariant conservative schemes for various shallow water models~\cite{dorodnitsyn2019shallow,bk:DorKapMelGN2020,DorKapSWJMP2021,bk:DorKapMel_ModSW_2021} and extended the Samarskiy--Popov schemes for the one-dimensional~MHD equations in plane geometry~\cite{bk:DorKapMDPI2022}.

In~\cite{bk:DorKozMelKap_CylFlows_2022}, the authors carried out a group classification of the~MHD equations in cylindrical geometry and obtained new conservation laws for both the case of finite and infinite conductivity.
The present publication is devoted to the application of these results to the case of finite-difference schemes for the~MHD equations.

\medskip

The paper is organized as follows.
In Section~\ref{sec:sec2}, the one-dimensional~MHD equations in cylindrical geometry, their symmetries and conservation laws for the case of finite conductivity are given.
Section~\ref{sec:sec3} is devoted to the construction of finite-difference schemes for the case of finite conductivity.
The obtained schemes possess conservation laws, including previously unknown ones.
The schemes are constructed on the basis of the classical results of~Samarskiy and~Popov.
A symmetry analysis of the derived schemes is carried out.
The case of infinite conductivity is considered in Section~\ref{sec:sec4}.
A new conservative scheme is again inherited from the scheme of a~Samarskiy--Popov type.
For the constructed scheme, it is possible to preserve two-point approximations of the entropy along the trajectories of motion.
Additional conservation laws obtained in~\cite{bk:DorKozMelKap_CylFlows_2022} are also considered. It is shown by an example that specific schemes can be constructed that also possess additional conservation laws.
In~Conclusion the obtained results are discussed.

\section{One-dimensional MHD flows with cylindrical symmetry}
\label{sec:sec2}

In the present section, the one-dimensional~MHD equations in mass~Lagrangian coordinates in cylindrical geometry in the case of finite conductivity, their symmetries, and conservation laws are considered.
Mass Lagrangian coordinates turn out to be especially convenient for setting boundary value problems in gas dynamics and magnetohydrodynamics~\cite{bk:SamarskyPopov_book[1992],bk:SamarskyPopov[1970],bk:YanenkRojd[1968]}, where a boundary arises naturally like boundary of a gas and vacuum etc.

\subsection{The MHD equations in mass Lagrangian coordinates}

The main system in mass Lagrangian coordinates~$(t,s)$  is
\begin{subequations}
\label{Lagrange_system}
\begin{gather}
\label{Lagrange_system_rho}
\rho_t = -\rho^2 (r u) _{s},
\\
\label{Lagrange_system_u}
u_t - \frac{v^2}{r}  = - r p_{s}
    - \frac{1}{2 r} \left( r^2 (H^{\theta})^2 \right)_s
    - \frac{r}{2} \left((H^z)^2\right)_s,
\\
\label{Lagrange_system_v}
v_t   +   \frac{u v}{r}
=  {H^r} \left(r H^{\theta}\right)_s,
\qquad
{\theta}_t  = \frac{v}{r},
\\
\label{Lagrange_system_w}
w_t = r {H^r} H^{z}_s,
\qquad
z_t = w,
\\
\label{Lagrange_system_p}
\varepsilon_t = -p (r u)_s + \frac{\sigma}{\rho} ((E^\theta)^2 + (E^z)^2),
\\
\label{Lagrange_system_H_theta}
H^{\theta}_t = r \rho ((v H^r + E^z)_s
    - H^{\theta} u_s ),
\\
\label{Lagrange_system_H_z}
H^z _t =  \rho (
    (r w H^r - r E^{\theta})_s
    - H^z (r u)_s),
\\
\label{Lagrange_system_relations}
\sigma E^{\theta} = -r \rho H^z_s,
\qquad
\sigma E^z = \rho (r H^{\theta})_s,
\end{gather}
\end{subequations}
where
\[
H^r = \frac{A}{r},
\qquad
A = \text{const},
\]
$t$ is time, $s$ is Lagrangian mass coordinate,
$\rho$ is density,
$p$ is pressure,
$\mathbf{u} = (u, v, w)$ is the velocity vector,
$\mathbf{E} = (0, E^\theta, E^z)$ and $\mathbf{H} = (H^r, H^\theta, H^z)$ are the electromagnetic field vectors,
$\sigma=\sigma(p, \rho)$ is conductivity,
$\varepsilon$ is internal energy,
$r,\theta, z$ are cylindrical coordinates.
We also notice that here and further~$f_t$ denotes the Lagrangian derivative of~$f$ with respect to time.

\smallskip

The Eulerian spatial coordinate~$r$ is a nonlocal variable
in the mass Lagrangian coordinates~\cite{bk:SamarskyPopov_book[1992],bk:YanenkRojd[1968]}:
\begin{subequations} \label{nonlocal_r}
\begin{equation} \label{nonlocal_r_t}
r_t = u,
\end{equation}
\begin{equation} \label{nonlocal_r_s}
r_s = \frac{1}{r\rho}.
\end{equation}
\end{subequations}

In~\cite{bk:SamarskyPopov_book[1992]} the following particular case $H^r=0$ is
considered~(in a slightly different form)
\begin{subequations}
\label{SysSamarskiy0}
\begin{gather}
\rho_t = -\rho^2 (r u) _{s},
\\
u_t = - r p_{s}
    - \frac{1}{2 r} \left( \kappa r^2 (H^{\theta})^2 \right)_s
    - \frac{r}{2} \left(\kappa (H^z)^2\right)_s,
\\
\varepsilon_t = -p (r u)_s + \frac{\sigma}{\rho} ((E^\theta)^2 + (E^z)^2),
\\
H^{\theta}_t = r \rho (E^z_s - H^{\theta} u_s ),
\\
H^z _t =  -\rho ( (r E^{\theta})_s + H^z (r u)_s),
\\
i^\theta = \sigma E^\theta = -\kappa\rho r H^z_s,
\qquad
i^z = \sigma E^z = \kappa\rho (r H^\theta)_s,
\end{gather}
\end{subequations}
where $\kappa = 1/(4\pi)$.

For systems (\ref{Lagrange_system}) and (\ref{SysSamarskiy0}) there are equivalence transformations~\cite{bk:DorKozMelKap_CylFlows_2022} that allow one to change the form of the equations while preserving their group properties. In particular, the equivalence transformation
\begin{equation}
\tilde{s} = \kappa s,
\quad
\tilde{p} = \kappa p,
\quad
\tilde{\rho} = \kappa \rho,
\quad
\tilde{\sigma} = \kappa \sigma
\end{equation}
allows one to put~$\kappa=1$.

In the sections devoted to finite-difference schemes, we will keep the factor~$\kappa$ in its original form,
since it may be essential in the numerical implementation of schemes~\cite{bk:DorKapMDPI2022}.

\medskip

Notice that the evolution equations for the magnetic field can be rewritten in the divergent form as
\begin{equation}
\left(\frac{H^\theta}{\rho}\right)_t = \frac{u H^\theta}{\rho r} + r E^z_s,
\qquad
\left(\frac{H^z}{\rho}\right)_t = -(r E^\theta)_s,
\end{equation}

\subsection{Symmetries of the MHD equations}
\label{sec:syms}

In the present section the symmetries of the MHD equations~(\ref{Lagrange_system_p}) are discussed
for the polytropic gas, i.e., when the equation of state~(\ref{InternalEnergyRel}) is held.
For the polytropic gas, by means of the equation of state
\begin{equation} \label{InternalEnergyRel}
\varepsilon = \frac{1}{\gamma - 1} \frac{p}{\rho},
\end{equation}
where $\gamma > 1$ is the polytrophic exponent,
equation~(\ref{Lagrange_system_p}) is brought to
\begin{equation}
p_t = -\gamma  \rho p (r u)_{s}
+  (\gamma - 1) \sigma  ((E^{\theta})^2  + (E^z)^2).
\end{equation}

Following the results derived in~\cite{bk:DorKozMelKap_CylFlows_2022}, we consider
two separate cases: $A \neq 0$ and $A = 0$.
It is also known from~\cite{bk:DorKozMelKap_CylFlows_2022} that the list of basic conservation laws corresponding to the arbitrary~$\sigma=\sigma(p, \rho)$ is extended by additional conservation laws only in case~$\sigma=C\rho$, $C\neq 0$.
Therefore, in the present section we consider only these two cases.

\begin{enumerate}[label=\arabic*)]

\item
\textbf{Case $A \neq 0$}

In case $\sigma$ is arbitrary (i.e., $\sigma=\sigma(p, \rho)$), the kernel of the admitted Lie algebras is given by
the generators
\begin{equation}     \label{kern01}
\def\arraystretch{2}
\begin{array}{c}
\displaystyle
X_1 = {\ddt},
\qquad
X_2 =  {\dds},
\qquad
X_{3} =  t  {\ddz} +  {\ddw}  ,
\\
\displaystyle
X_4 =  f_1 (s) { \partial \over \partial \theta} ,
\qquad
X_{5} = f _2 (s)   {\ddz},
\end{array}
\end{equation}
where $f _1(s) $  and  $f _2(s) $  are arbitrary functions of their arguments.

In case $\sigma=C\rho$, the extension of the kernel of the admitted Lie algebra consists of the generator
\begin{multline} \label{kern01_ext1}
X_6 = 2 t {\ddt} + 2 s {\dds} - u  {\ddu} - v  {\ddv} - w  {\ddw}
    + r  {\ddr}   + z  {\ddz}  - 2 p {\ddp}
    \\
    - 2 E^{\theta} {\ddEtheta} - 2 E^{z}  {\ddEz}
    -  H^{\theta}  {\ddHtheta} -  H^z  {\ddHz}.
\end{multline}

\item
\textbf{Case $A = 0$}

In case $\sigma$ is arbitrary, the kernel of the admitted Lie algebras is
\begin{equation}     \label{kern02}
X_1^0 = {\ddt},
\qquad
X_2^0 =  {\dds},
\qquad
X_3^0 =  f( s, r v ) { \partial \over \partial \theta} ,
\end{equation}
where the function  $f( s, r v )$ is arbitrary.

In case $\sigma=C\rho$, the extension consists of two generators, namely
\begin{multline} \label{kern02_ext1}
X_4^0 = 2 t   {\ddt} + 2 s   {\dds}
    - u  {\ddu} - v  {\ddv}
    + r  {\ddr}
    -  2 p  {\ddp}
    \\
    - 2 E^{\theta}   {\ddEtheta}
    -  2 E^{z}   {\ddEz}
    -  H^{\theta}  {\ddHtheta}  -  H^z  {\ddHz},
\end{multline}
\begin{multline} \label{kern02_ext2}
X_5^0 = 2 t {\ddt} + 2 s {\dds}
- 2 u  {\ddu} - 2 v  {\ddv}
 - 2 p   {\ddp} + 2 \rho  {\ddrho}
 \\
    - 3 E^{\theta}   {\ddEtheta} - 3 E^z   {\ddEz}
    - H^{\theta}   {\ddHtheta} - H^z  {\ddHz}.
\end{multline}

\end{enumerate}

\subsection{Conservation laws of the MHD equations}

The conservation laws for the system of equations~(\ref{nonlocal_r}), (\ref{Lagrange_system}), which were obtained in~\cite{bk:DorKozMelKap_CylFlows_2022}, are listed in the following sub sections.
Some of these conservation laws have not been previously known.

\subsubsection{Case $A \neq 0$}

The conservation laws are the following
\begin{itemize}
\item

mass
\begin{equation}   \label{CL_general_A_mass}
D_t
\left(
{1 \over \rho }
\right)
-
D_{s}
\left(
{ r u  }
\right)
= 0  ;
\end{equation}

\item

momentum along $z$-axis
\begin{equation}   \label{CL_general_A_momentum_z}
D_t
\left(
   w
\right)
-
D_{s}
\left(
 r H^r   H^{z}
\right)
= 0  ;
\end{equation}

\item

motion of the center of mass  along $z$-axis
\begin{equation}    \label{CL_general_A_galileo_z}
D_t
\left(
 t w   - z
\right)
-
D_{s}
\left(
   t r H^r    H^{z}
\right)
= 0 ;
\end{equation}

\item

angular momentum in $ (r, \theta )$-plane
\begin{equation}   \label{CL_general_A_rotation}
D_t
\left(
    r v
\right)
-
D_{s}
\left(
  r ^2 H^r   H^{\theta}
\right)
= 0  ;
\end{equation}

\item

magnetic fluxes
\begin{equation}            \label{CL_general_A_flux_theta}
D_t
\left(
H^{\theta}   \over  r \rho
\right)
-
D_{s}
\left(
E^z   +
 v H^r
\right)
= 0  ,
\end{equation}
\begin{equation}           \label{CL_general_A_flux_z}
D_t
  \left(
H^z   \over  \rho
\right)
+
D_{s}
\left(
 r E ^{\theta}
-
r w H^r
\right)
= 0    ;
\end{equation}

\item

energy
\begin{multline}   \label{CL_general_A_energy}
D_t
\left\{
{  1 \over  \gamma  -1  } { p \over    \rho  }
+ { 1   \over 2 }    (    u  ^2 + v ^2 +  w^2  )
+
 {   ( H^{\theta}  ) ^2  +  ( H^{z}  ) ^2   \over  2  \rho }
\right\} \\
+
D_{s}
\left\{
      r  u  \left(
  p
  +
    {  ( H^{\theta}  ) ^2  +  ( H^{z}  ) ^2   \over 2}
 \right)
+ r ( E ^{\theta}  H ^z  - E ^z H ^{\theta} )
 -  r H^r  ( v H^{\theta} + w H^{z}  )
\right\}
=  0  .
\end{multline}
\end{itemize}
For electric conductivity $\sigma ( \rho , p)  =  C \rho $ with constant   $ C \neq 0  $
there exists the additional conservation law
\begin{equation}      \label{special_CL_1}
D_t \left\{
    \left(2 t - C s \right) \frac{H^z}{\rho} - C r z H^r
\right\}
+ D_s \left\{
    \left( 2 t - C s \right) (r E^{\theta} - r w H^r  )
    - r^2 H^z
\right\} = 0.
\end{equation}

\subsubsection{Case $A = 0$}

\begin{itemize}
\item

mass
\begin{equation}   \label{CL_A0_mass}
D_t
\left(
{1 \over \rho }
\right)
-
D_{s}
\left(
{ r u  }
\right)
= 0  ;
\end{equation}

\item

angular momentum in $ (r, \theta )$-plane
\begin{equation}   \label{CL_A0_rotation}
D_t
\left(
    r v
\right)
= 0  ;
\end{equation}

\item

magnetic  fluxes
\begin{equation}            \label{CL_A0_flux_theta}
D_t
\left(
H^{\theta}   \over  r \rho
\right)
-
D_{s}
\left(
E^z
\right)
=  0   ,
\end{equation}
\begin{equation}          \label{CL_A0_flux_z}
D_t
  \left(
H^z   \over  \rho
\right)
+
D_{s}
\left(    r E ^{\theta}
\right)
= 0   ;
\end{equation}

\item

energy
\begin{multline}    \label{CL_A0_energy}
D_t
\left\{
{  1 \over  \gamma  -1  } { p \over    \rho  }
+ { 1   \over 2 }    (    u  ^2 + v ^2  )
+  {   ( H^{\theta}  ) ^2  +  ( H^{z}  ) ^2   \over  2  \rho }
\right\}
\\
+
D_{s}
\left\{
      r  u  \left(
  p
  +
    {  ( H^{\theta}  ) ^2  +  ( H^{z}  ) ^2   \over 2}
 \right)
+ r ( E ^{\theta}  H ^z  - E ^z H ^{\theta} )
\right\}
=  0  .
\end{multline}
\end{itemize}
For conductivity   $\sigma ( \rho , p) =  C \rho $, $ C \neq 0,$
there are two additional conservation laws
\begin{equation}    \label{special_CL_2}
D_t  \left\{
    \left( 2 t - C s \right) \frac{H^z}{\rho}
\right\}
+ D_s\left\{
    \left( 2 t - C s \right) r E^{\theta}
    - r^2 H^z
\right\}= 0
\end{equation}
and
\begin{equation}    \label{special_CL_3}
D_t \left(
    C  s   \frac{H^\theta}{r \rho}
\right)
- D_s\left(
    C s E^z
    - r H^\theta
\right) = 0.
\end{equation}
Notice that the conservation laws (\ref{special_CL_1}), (\ref{special_CL_2}), and (\ref{special_CL_3}) obtained in~\cite{bk:DorKozMelKap_CylFlows_2022}, to the best of the authors' knowledge, have not been known before.

\subsubsection{Various forms of the conservation law of energy}
\label{sec:forms_of_energy_cls}

Here we consider different forms of the conservation law of energy~(\ref{CL_general_A_energy}).
Being equivalent in the mathematical sense, they reflect different physical aspects of the phenomenon.
In numerical modeling, it makes sense to construct \emph{completely} conservative schemes~\cite{bk:SamarskyPopov_book[1992]} that not only have finite-difference analogues of conservation laws, but also correctly hold the balance between internal energy, gas-dynamic energy, and total energy.

\smallskip

Based on the results given in~\cite{bk:SamarskyPopov_book[1992]}, one obtains the following non-divergence forms of the conservation law of energy~(\ref{CL_general_A_energy}) for system~(\ref{Lagrange_system}), (\ref{nonlocal_r})
\begin{equation} \label{CL_general_A_energy_V2}
\varepsilon_t + p \left(r u\right)_s - q = 0,
\end{equation}
\begin{equation} \label{CL_general_A_energy_V3}
D_t\left(\varepsilon + \frac{u^2 + v^2 + w^2}{2}\right)
    + \left(
        r p u
    \right)_s
    - u f^r
    - v f^\theta
    - w f^z
    - q = 0,
\end{equation}
where the magnetic force $(f^r, f^\theta, f^z)$ and the Joule heating~$q$ per unit mass are given by
\begin{equation}
\def\arraystretch{2}
\begin{array}{c}
\displaystyle
f^r = \frac{\sigma (E^\theta H^z - E^z H^\theta)}{\rho},
\qquad
f^\theta = H^r (r H^y)_s,
\qquad
f^z = r H^r H^z_s,
\\
\displaystyle
q = \frac{\sigma \left((E^\theta)^2 + (E^z)^2\right)}{\rho},
\end{array}
\end{equation}
and the quantity $p \left(r u\right)_s$ characterizes the work of the gas-kinetic pressure forces.

Recall that the conservation law, written in divergent form~(\ref{CL_general_A_energy}), shows that the total energy (internal, kinetic and magnetic) changes due to the work of the forces of gas-kinetic and magnetic pressure and the flow of electromagnetic energy.
Non-divergent-form conservation laws~(\ref{CL_general_A_energy_V2}) and~(\ref{CL_general_A_energy_V3}) respectively describe the evolution of internal energy and gas-dynamic energy (i.e., internal and kinetic energy).


\section{Conservative schemes for the MHD equations}
\label{sec:sec3}

Further consideration is based on the results derived in~\cite{bk:SamarskyPopov[1970],bk:SamarskyPopov_book[1992]}, where completely conservative finite-difference schemes for the~MHD equations for cylindrical flows were constructed. These schemes were constructed for the system of equations~(\ref{nonlocal_r}), (\ref{SysSamarskiy0}), in which the angular and axial components of the velocity and the radial component of the magnetic field were discarded.
Further we consider these schemes in more detail, extend them to a more general case, and study their symmetries and conservation laws.

\medskip

In the following sections, the specific notation is used to shorten the representation of the finite-difference expressions.
For brevity, a value of difference function $f$ of two variables at point~$(t_n, s_m)$ is denoted as~$f^n_m$.
Finite-difference derivatives of some quantity~$\phi=\phi(t_n, s_m, u^n_m, ...)$ with respect to the variables~$t$ and~$s$ are denoted as
\begin{equation}
\def\arraystretch{2.0}
\begin{array}{c}
\displaystyle
\phi_t = \underset{+\tau}{D}(\phi)
    = \frac{\dtp{S}(\phi) - \phi}{\tau_n},
\qquad
\phi_s = \underset{+s}{D}(\phi)
    = \frac{\dsp{S}(\phi) - \phi}{h_m},
\\
\displaystyle
\phi_{\check{t}} = \underset{-\tau}{D}(\phi)
    = \frac{\phi - \dtm{S}(\phi)}{\tau_{n-1}},
\qquad
\phi_{\bar{s}} = \underset{-s}{D}(\phi)
    = \frac{\phi - \dsm{S}(\phi)}{h_{m-1}},
\end{array}
\end{equation}
where $\underset{\pm\tau}{D}$ and $\underset{\pm h}{D}$ are finite-difference total differentiation operators.
They are defined through the  left and right finite-difference shifts along the time and space (or mass)~$s$ axes
\[
    \def\arraystretch{1.75}
    \begin{array}{c}
    \displaystyle
    \underset{\pm\tau}{S}(\phi(t_n, s_m, u^{n}_{m}, ...))
        = \phi(t_{n \pm 1}, s_m, u^{n \pm 1}_{m}, ...),
    \\
    \displaystyle
    \underset{\pm{s}}{S}(\phi(t_n, s_m, u^{n}_{m}, ...))
        = \phi(t_n, s_{m\pm 1}, u^{n}_{m \pm 1}, ...).
    \end{array}
\]
The indices~$n$ and~$m$ respectively change along the axes~$t$
and~$s$. The finite-difference mesh steps $h_m$, $\tau_n$ are defined as
\begin{equation}
\def\arraystretch{1.25}
\begin{array}{c}
\displaystyle
\tau_n = t_{n+1} - t_n,
\qquad
\tau_{n-1} = t_n - t_{n-1},
\\
\displaystyle
h_m = s_{m+1} - s_m,
\qquad
h_{m-1} = s_{m} - s_{m-1}.
\end{array}
\end{equation}
Further consideration is restricted to uniform meshes for which
\begin{equation}
h_m = h_{m-1} = h = \text{const},
\qquad
\tau_n = \tau_{n-1} = \tau = \text{const},
\qquad
m, n \in \mathbb{Z}.
\end{equation}

\medskip

Following the Samarskiy--Popov notation throughout the text, we denote
\begin{equation}
\dsp{S}(\phi) = \phi_+,
\quad
\dsm{S}(\phi) = \phi_-,
\quad
\dtp{S}(\phi) = \hat{\phi},
\quad
\dtm{S}(\phi) = \check{\phi},
\end{equation}

\begin{equation}
{\phi}^{(\alpha)}=\alpha\hat{{\phi}}+(1-\alpha){\phi},
\end{equation}
\begin{equation}
\label{eq:star}
{\phi}^* \equiv {\phi}_*=({\phi}_*)^j_i=\frac{h_i {\phi}^j_{i-1/2}+h_{i-1}{\phi}^j_{i+1/2}}{h_i+h_{i-1}}.
\end{equation}
In particular,
\begin{equation}
\phi^{(0.5)} = \frac{\hat{\phi} + {\phi}}{2}.
\end{equation}
Notice that in case $h_m = h = \text{const}$ in its integral nodes~(\ref{eq:star}) is just
\begin{equation}
\displaystyle
{\phi}_* = \frac{{\phi}_- + {\phi}}{2}.
\end{equation}

\bigskip

In what follows, we consider the case of finite conductivity~$\sigma(p, \rho)$.
The results for the case of infinite conductivity ($\sigma \to \infty$), on one hand, can be partially inherited from the case of finite conductivity.
On the other hand, in case~$\sigma \to \infty$, a more complex group classification arises, leading to the occurrence  of a large number of additional conservation laws, which requires its separate consideration.
This is the subject of Section~\ref{sec:inf}.

\subsection{Schemes for the case of finite conductivity}

\subsubsection{The classical Samarskiy--Popov scheme}

The conservative scheme for system~(\ref{nonlocal_r}), (\ref{SysSamarskiy0}) is
\begin{subequations}
\label{SamPopScheme1}
\begin{gather} \label{SamPopScheme1_mass}
\left(\frac{1}{\rho}\right)_t = (r^{(0.5)} u^{(0.5)})_s,
\\
u_t = -r^{(0.5)} p^{(\alpha)}_{\bar{s}} + f^r,
\qquad
r_t = u^{(0.5)},
\\ \label{SamPopScheme1_fr}
f^r = -\kappa r^{(0.5)} \left(\frac{H\hat{H}}{2}\right)_{\bar{s}}
    - \frac{\kappa}{r^{(0.5)}} \left( {b} \frac{h^\theta\hat{h}^\theta}{2} \right)_{\bar{s}},
\\
\left(\frac{H}{\rho}\right)_t = -e^{(\lambda)}_s,
\\
\left(\frac{h^\theta}{\rho a^2}\right)_t = E_s^{(\beta)},
\\
I = -\kappa \rho_* r H_{\bar{s}} = \sigma_* \frac{e}{r},
\\
i = \kappa \rho_* (h^\theta)_{\bar{s}} = \sigma_* E,
\\ \label{SamPopScheme1_energy}
\varepsilon_t = -p^{(\alpha)} (r^{(0.5)} u^{(0.5)})_s + q,
\\ \label{SamPopScheme1_q}
q = \left[
    \left(\frac{I_+}{\rho^*_+ r_+}\right)^{(0.5)} e^{(\lambda)}_+
    + \left(\frac{i_+}{\rho^*_+}\right)^{(0.5)} E^{(\beta)}_+
\right]_*,
\end{gather}
\end{subequations}
where
\begin{equation} \label{SamPopSchemeAbbr}
\def\arraystretch{1.75}
\begin{array}{c}
H = (H^z)^j_{i+1/2},
\qquad
h^\theta = (r H^\theta)^j_{i+1/2},
\qquad
E = (E^z)^j_i,
\\
{\mathcal{E}} = (r E^\theta)^j_i,
\qquad
{b} = r^{(0.5)} r^{(0.5)}_+ / (r_* \hat{r}_*)_+,
\end{array}
\end{equation}
and the weighting coefficients
\[
\alpha, \beta, \lambda \in [0, 1]
\]
determine the distribution of the quantities $\phi^{(\alpha)}$, $\phi^{(\beta)}$, and $\phi^{(\lambda)}$ over time layers.
For any values of~$\alpha$, $\beta$, and~$\lambda$ the scheme is of the first order of approximation by~$\tau$ and~$h$.

\begin{remark}
\label{rem:r_s}
In~\cite{bk:SamarskyPopov_book[1992]}, no approximation of equation~(\ref{nonlocal_r_s}) is explicitly given.
Analyzing the conservation laws of scheme~(\ref{SamPopScheme1}), one establishes that the desired approximation has the form
\begin{equation} \label{SamPop_r_s_approx}
r_s = \frac{2}{(r + r_+) \rho}.
\end{equation}
Thus, further the system of equations~(\ref{SamPopScheme1}) together with equation~(\ref{SamPop_r_s_approx}) is considered.
\end{remark}

According to~\cite{bk:SamarskyPopov_book[1992]}, scheme~(\ref{SamPopScheme1}) possesses the following conservation laws
\begin{itemize}
    \item
    mass
    \begin{equation}
    \left(\frac{1}{\rho}\right)_t - (r^{(0.5)} u^{(0.5)})_s = 0;
    \end{equation}

    \item
    magnetic flux
    \begin{equation} \label{SamPopScheme1_CL_Hz}
    \left(\frac{H}{\rho}\right)_t + ({\mathcal{E}}^{(\lambda)})_s = 0,
    \end{equation}
    \begin{equation} \label{SamPopScheme1_CL_Hy}
    \left(\frac{h^\theta}{\rho (r^*_+)^2}\right)_t - (E^{(\beta)})_s = 0;
    \end{equation}

    \item
    total energy
    \begin{multline} \label{SamPopScheme1_energy_DIV}
        \left(
            \varepsilon
            + \frac{u^2 + u_+^2}{4}
            + \frac{\kappa H^2}{2 \rho}
            + \frac{\kappa (h^\theta)^2}{2 \rho (r^*_+)^2}
        \right)_t
        + \left[
            \left(p_*^{(\alpha)} + \frac{\kappa (H\hat{H})_*}{2}\right) r^{(0.5)} u^{(0.5)}
            \right.
            \\
            \left.
            + \frac{\kappa}{2} \frac{u^{(0.5)}}{r^{(0.5)}} \left({b} h^\theta \hat{h}^\theta \right)_*
            + \kappa \left({\mathcal{E}}^{(\lambda)} H_*^{(0.5)} - E^{(\beta)} (h^{\theta})^{(0.5)}_*\right)
        \right]_s = 0;
    \end{multline}
\end{itemize}

Similar to the continuous case, the scheme possesses additional conservation laws of a special form in case~$\sigma = C \rho$.
Indeed, using (\ref{SamPopScheme1_CL_Hz}), (\ref{SamPopScheme1_CL_Hy}), and (\ref{SamPop_r_s_approx}), performing standard algebraic calculations, one derives two additional conservation laws
\begin{itemize}
\item
the finite-difference analogue of~(\ref{special_CL_2})
\begin{equation}
\displaystyle
\left[
    \left(2 \check{t} - \frac{C s}{\kappa}\right) \frac{H}{\rho}
\right]_t
+ \left[
    \left(2 (t - \lambda\tau) - \frac{C s_-}{\kappa} \right) {\mathcal{E}}^{(\lambda)}
    - (r^2 H_-)^{(\lambda)}
\right]_s = 0;
\end{equation}

\item
the finite-difference analogue of~(\ref{special_CL_3})
\begin{equation}
\left(
    \frac{C s}{\kappa} \frac{h^\theta}{\rho (r^*_+)^2}
\right)_t
- \left(
    \frac{C s_-}{\kappa} E^{(\beta)}
    - (h^\theta_-)^{(\beta)}
\right)_s = 0.
\end{equation}

\end{itemize}

\begin{remark}
\label{rem:evol_H_sq}
One can verify that the magnetic flux conservation laws~(\ref{SamPopScheme1_CL_Hz}) and~(\ref{SamPopScheme1_CL_Hy}) are equivalent to the following equations, which are useful for constructing various forms of the conservation law of energy
\begin{equation}
\left(\frac{H^2}{\rho}\right)_t + (\hat{H} + H) {\mathcal{E}}^{(\lambda)}_s + \hat{H} H \left(\frac{1}{\rho}\right)_t = 0,
\end{equation}
\begin{equation}
\left(\frac{(h^\theta)^2}{\rho (r^*_+)^2}\right)_t - (\hat{h}^\theta + h^\theta) E^{(\beta)}_s + \hat{h}^\theta h^\theta \left(\frac{1}{\rho (r^*_+)^2}\right)_t = 0.
\end{equation}
Notice that the latter equations by their own have the physical meaning of the evolution of the axial and angular components of the magnetic pressure.
\end{remark}

\begin{remark}
As in the differential case, there are three different forms of the conservation law of energy: equations~(\ref{SamPopScheme1_energy}) and~(\ref{SamPopScheme1_energy_DIV}) approximate~(\ref{CL_general_A_energy_V2}) and~(\ref{CL_A0_energy}), and the approximation for~(\ref{CL_general_A_energy_V3}) is
\begin{equation}
    \left(\varepsilon + \frac{u^2 + u_+^2}{4}\right)_t
    + (p^{(\alpha)}_* r^{(0.5)} u^{(0.5)})_s
    - \frac{1}{2}((u_+ + \hat{u}_+) f^r_+)_*
    - q = 0,
\end{equation}
where $f^r$ and $q$ are given by (\ref{SamPopScheme1_fr}) and (\ref{SamPopScheme1_q}).

Thus, the balance between various forms of energy is also held in the finite-difference case (i.e., the scheme is completely conservative). In particular, the latter non-divergent form of the conservation law describes
the evolution of the gas-dynamic energy. Recall that the physical meaning of the various forms of the energy conservation law
was discussed in Section~\ref{sec:forms_of_energy_cls}.
\end{remark}

\begin{remark}
To preserve the angular momentum in $ (r, \theta )$-plane, one should extend the scheme
with an evolution equation for the angular component~$v$ of the velocity.
For example, one can choose the following approximation for the first equation of~(\ref{Lagrange_system_v})
\begin{equation} \label{scheme1_CL_angmom}
v_t + \frac{\hat{v} u^{(0.5)}}{r} = 0.
\end{equation}
In this case, taking into account~(\ref{SamPopScheme1_mass}), the finite-difference analogue of (\ref{CL_A0_rotation}) appears as
\begin{equation} \label{angularMomCL_triv}
\left(r v\right)_t = \frac{1}{\tau}(\hat{r}\hat{v} - r v) = 0,
\end{equation}
i.e., in absence of the radial component of the magnetic field, the quantity $r v$ is preserved along the trajectories of motion.
\end{remark}

\subsubsection{Extension of the Samarskiy--Popov scheme}

Scheme (\ref{SamPopScheme1}) can be extended for the case of a non-zero radial component $H^r$ of the magnetic field, i.e., $H^r = \frac{A}{r}$, $A \neq 0$. The main difference from the case $H^r=0$ is the presence of the angular and axial velocity components,~$v$ and~$w$.

\medskip

Since we are interested in schemes with conservation laws preservation, we extend scheme~(\ref{SamPopScheme1}) in such a way that it still possesses the largest possible set of conservation laws.

The scheme extension procedure involves a large number of calculations, so here we only discuss in detail how the equations can be extended to preserve the conservation laws of energy end angular momentum.
First, we recall~\cite{bk:Dorodnitsyn[2011]} that, by analogy with differential equations, any finite-difference
conservation law of a system of~$N$ difference equations
\[
F^j = 0, \qquad j = 1, 2, \dots, N,
\]
can be represented as a sum
\[
\displaystyle \sum_{j=1}^N \Lambda_j F^j = 0,
\]
where the quantities $\Lambda_j$, $j=1, \dots, N,$ are called finite-difference conservation law multipliers.
In particular, in case $H^r=0$ the conservation law of energy~(\ref{SamPopScheme1_energy_DIV})
can be rewritten in terms of equations~(\ref{SamPopScheme1}) and multipliers as
\begin{multline} \label{SamPopEnergyCL_detailed}
    \left(
        \varepsilon
        + \frac{u^2 + u_+^2}{4}
        + \frac{\kappa H^2}{2 \rho}
        + \frac{\kappa (h^\theta)^2}{2 \rho (r^*_+)^2}
    \right)_t
    + \left[
            \left(p_*^{(\alpha)} + \frac{\kappa (H\hat{H})_*}{2}\right) r^{(0.5)} u^{(0.5)}
            \right.
            \\
            \left.
            + \frac{\kappa}{2} \frac{u^{(0.5)}}{r^{(0.5)}} \left({b} h^\theta \hat{h}^\theta \right)_*
            + \kappa \left({\mathcal{E}}^{(\lambda)} H_*^{(0.5)} - E^{(\beta)} (h^{\theta})^{(0.5)}_*\right)
        \right]_s
    \\
    =
    \Lambda_1 \cdot (\varepsilon_t + p^{(\alpha)} (r^{(0.5)} u^{(0.5)})_s - q) \\
    + \Lambda_2 \cdot (u_t + r^{(0.5)} p^{(\alpha)}_{\bar{s}} - f^r)
    + \Lambda_3 \cdot (u_t^+ + r^{(0.5)}_+ p^{(\alpha)}_{s} - f^r_+)
    \\
    + \Lambda_4 \cdot \left\{
        \left(\frac{H^2}{\rho}\right)_t + (\hat{H} + H) {\mathcal{E}}^{(\lambda)}_s + \hat{H} H \left(\frac{1}{\rho}\right)_t
    \right\}
    \\
    + \Lambda_5 \cdot \left\{
        \left(\frac{(h^\theta)^2}{\rho (r^*_+)^2}\right)_t - (\hat{h}^\theta + h^\theta) E^{(\beta)}_s
        + \hat{h}^\theta h^\theta \left(\frac{1}{\rho (r^*_+)^2}\right)_t
    \right\},
\end{multline}
where
\begin{equation}
\Lambda_1 = 1,
\qquad
\Lambda_2 = \frac{u + \hat{u}}{4},
\qquad
\Lambda_3 = \Lambda_2^+ = \frac{u_+ + \hat{u}_+}{4},
\qquad
\Lambda_4 = \Lambda_5 = \frac{\kappa}{2}
\end{equation}
are the finite-difference conservation law multipliers.
Such representations of conservation laws were considered in~\cite{bk:SamarskyPopov_book[1992]} when constructing completely conservative schemes for the equations of gas dynamics and magnetohydrodynamics.
Equation~(\ref{SamPopEnergyCL_detailed}) was apparently also derived by the authors of~\cite{bk:SamarskyPopov_book[1992]}, although it was not explicitly given by them.

\smallskip

Now we extend the described results to the case~$H^r \neq 0$.
According to system~(\ref{Lagrange_system}) and the conservation law~(\ref{CL_general_A_energy}), the conservation law~(\ref{SamPopEnergyCL_detailed}) should be extended as
\begin{multline} \label{approx0_for_cl_energy}
    \left(
        \varepsilon
        + \frac{u^2 + u_+^2}{4}
        + \frac{\kappa H^2}{2 \rho}
        + \frac{\kappa (h^\theta)^2}{2 \rho (r^*_+)^2}
        + A_1
    \right)_t + \left(\cdots\right)_s
    =
    \Lambda_1 \cdot (\varepsilon_t + p^{(\alpha)} (r^{(0.5)} u^{(0.5)})_s - q) \\
    + \Lambda_2 \cdot (u_t - A_2 + r^{(0.5)} p^{(\alpha)}_{\bar{s}} - f^r)
    + \Lambda_3 \cdot (u_t^+ - A_2^+ + r^{(0.5)}_+ p^{(\alpha)}_{s} - f^r_+)
    \\
    + \cdots
    + \Lambda_6 \cdot (v_t + A_3 + \cdots) + \cdots + D_1,
\end{multline}
where $A_1$, $A_2$, $A_3$, and $\Lambda_6$ are some approximations
\begin{equation}
A_1 \sim \frac{v^2}{r},
\qquad
A_2 \sim \frac{v^2}{r},
\qquad
A_3 \sim \frac{u v}{r},
\qquad
\Lambda_6 \sim v,
\end{equation}
corresponding to the additional terms that arise due to~$H^r \neq 0$,
$D_1$ is some finite-difference divergent expression of order $O(h)$,
and~`$\cdots$' denote the remaining terms of~(\ref{SamPopEnergyCL_detailed}).

Taking into account~(\ref{scheme1_CL_angmom}), we choose
\begin{equation}
A_3 = \frac{\hat{v} u^{(0.5)}}{r}.
\end{equation}
This allows us to hold the form of the conservation law of angular momentum~(\ref{angularMomCL_triv}) in case~$A\neq 0$.

Then, due to~(\ref{approx0_for_cl_energy}), our choice of approximation~$A_3$ imposes the following restrictions
on~$A_1$, $A_2$ and~$\Lambda_6$
\begin{equation}
\frac{\hat{A}_1 - A_1}{\tau}
    + \frac{u + \hat{u}}{4} A_2
    + \frac{u_+ + \hat{u}_+}{4} A_2^+
    - \Lambda_6 \cdot \frac{\hat{v} u^{(0.5)}}{r} = D_1.
\end{equation}

Considering various possible approximations of a general form by introducing
indeterminate coefficients, we find that the latter restriction is satisfied by
\begin{align}
A_1 = \frac{v^2}{2},
\qquad
A_2 = \frac{\hat{v} v^{(0.5)}}{r},
\qquad
\Lambda_6 = \frac{v + \hat{v}}{2},
\\
D_1 = \left(\frac{h}{2\tau}\left(\frac{\hat{r}}{r}-1\right) \hat{v} v^{(0.5)}\right)_s
= \left(\frac{h \, \hat{v} v^{(0.5)} r_t}{2 r}\right)_s.
\end{align}
Turning back to~(\ref{approx0_for_cl_energy}), one sees that,
by studying two representations of the conservation law of total energy, we established the desired form of the equations of the extended finite-difference scheme.
Namely, we have established the terms~$A_2$ and~$A_3$ involved in the extension of the evolution equations
for the velocity components~$u$ and~$v$.

The remaining equations of scheme~(\ref{SamPopScheme1}) are extended in a similar way.
For the equations that can be extended in more than one way, we choose the simplest approximations.
As a final result we derive the scheme
\begin{subequations} \label{SamPopSchemeExt1}
\begin{equation}
\left(\frac{1}{\rho}\right)_t = (r^{(0.5)} u^{(0.5)})_s,
\end{equation}
\begin{equation} \label{SamPopSchemeExt1_u}
u_t - \frac{\hat{v} v^{(0.5)}}{r} + r^{(0.5)} p^{(\alpha)}_{\bar{s}} - f^r = 0,
\end{equation}
\begin{equation} \label{SamPopSchemeExt1_v}
v_t + \frac{\hat{v} u^{(0.5)}}{r}
    - \kappa H^r (h^\theta)^{(0.5)}_s = 0,
\end{equation}
\begin{equation}
w_t - \kappa \left( r H^r H^{(0.5)} \right)_s = 0,
\end{equation}
\begin{equation} \label{SamPopScheme1_CL_Hz_ext}
\left(\frac{H}{\rho}\right)_t
+ {\mathcal{E}}^{(\lambda)}_s
    - (r H^r w^{(0.5)})_{\bar{s}}
= 0,
\end{equation}
\begin{equation} \label{SamPopScheme1_CL_Hy_ext}
\left(\frac{h^\theta}{\rho (r^*_+)^2}\right)_t -
        E^{(\beta)}_s
        - (H^r v^{(0.5)})_{\bar{s}}
= 0,
\end{equation}
\begin{equation} \label{SamPopScheme1_energy_evol}
\varepsilon_t = -p^{(\alpha)} (r^{(0.5)} u^{(0.5)})_s + q,
\end{equation}
\begin{equation} \label{SamPopScheme1_coord_rels}
r_t = u^{(0.5)},
\qquad
r_s = \frac{2}{(r + r_+) \rho},
\end{equation}
\begin{equation}
z_t = w^{(0.5)},
\qquad
\theta_t = \frac{v^{(0.5)}}{r^{(0.5)}},
\end{equation}
\end{subequations}
where $f^r$ and $q$ are given by (\ref{SamPopScheme1_fr}) and (\ref{SamPopScheme1_q}).

As it was discussed in~Remark~\ref{rem:r_s}, equations~(\ref{SamPopScheme1_coord_rels})
are included in the scheme. They relate the~Eulerian spatial coordinate~$r$ with the mass~Lagrangian
coordinate~$s$. These equations are also necessary for some of the
conservation laws to be satisfied on solutions of the scheme.

\bigskip

The construction procedure of scheme~(\ref{SamPopSchemeExt1}) guarantees the scheme to possess the following conservation laws:
\begin{itemize}
\item
mass
\begin{equation}
\left(\frac{1}{\rho}\right)_t = (r^{(0.5)} u^{(0.5)})_s;
\end{equation}

\item
center-of-mass law
\begin{equation}
\left(t w^{(0.5)} - z\right)_t
- \left( \kappa \hat{t} \left(r H^r H^{(0.5)}\right)^{(0.5)} \right)_s = 0;
\end{equation}

\item
magnetic fluxes
\begin{equation}
\left(\frac{H}{\rho}\right)_t
+ \left(
    {\mathcal{E}}^{(\lambda)}
    - (r H^r w^{(0.5)})_-
\right)_s = 0,
\end{equation}
\begin{equation}
\left(\frac{h^\theta}{\rho (r^*_+)^2}\right)_t -
    \left(
        E^{(\beta)}
        + (H^r v^{(0.5)})_-
    \right)_s = 0;
\end{equation}

\item
angular momentum in $(r, \theta )$-plane
\begin{equation} \label{SchemeCL_angmom}
(r v)_t - (\kappa r H^r (h^\theta)^{(0.5)})_s = 0;
\end{equation}

\item
total energy
\begin{multline} \label{SamPopSchemeExt_energy_DIV}
        \left(
            \varepsilon
            + \frac{u^2 + u_+^2}{4}
            + \frac{v^2 + w^2}{2}
            + \frac{\kappa H^2}{2 \rho}
            + \frac{\kappa (h^\theta)^2}{2 \rho (r^*_+)^2}
        \right)_t
        + \left[
            \left(p_*^{(\alpha)} + \frac{\kappa (H\hat{H})_*}{2}\right) r^{(0.5)} u^{(0.5)}
            \right.
            \\
            \left.
            + \frac{\kappa}{2} \frac{u^{(0.5)}}{r^{(0.5)}} \left({b} h^\theta \hat{h}^\theta \right)_*
            + \kappa \left({\mathcal{E}}^{(\lambda)} H_*^{(0.5)} - E^{(\beta)} (h^{\theta})^{(0.5)}_*\right)
            -\frac{h}{2\tau}\left(\frac{\hat{r}}{r}-1\right) \hat{v} v^{(0.5)}
            \right.
            \\
            \left.
            - \kappa (h^\theta)^{(0.5)} (H^r v^{(0.5)})_-
            - \kappa r H^r w_-^{(0.5)} H^{(0.5)}
        \right]_s = 0;
 \end{multline}
Notice that the scheme possesses alternative, non-divergent, forms of the conservation law of energy~(i.e., the scheme is completely conservative):
\begin{itemize}[label=$\circ$]
\item
the finite-difference analogue of (\ref{CL_general_A_energy_V2}) which describes the evolution of internal energy
\begin{equation}
\varepsilon_t = -p^{(\alpha)} (r^{(0.5)} u^{(0.5)})_s + q,
\end{equation}
where $q$ is given by (\ref{SamPopScheme1_q});

\item
analogue of (\ref{CL_general_A_energy_V3}) which describes the evolution of gas-dynamic energy
\begin{multline}
    \left(\varepsilon + \frac{u^2 + u_+^2}{4} + \frac{v^2 + w^2}{2}\right)_t
    + \left(
        p^{(\alpha)}_* r^{(0.5)} u^{(0.5)}
        -\frac{h}{2\tau}\left(\frac{\hat{r}}{r}-1\right) \hat{v} v^{(0.5)}
    \right)_s
    \\
    - (u^{(0.5)}_+ f^r_+)_*
    - v^{(0.5)} f^\theta
    - w^{(0.5)} f^z
    - q = 0,
\end{multline}
where
\begin{equation}
f^\theta = \kappa H^r (h^\theta)^{(0.5)}_s,
\qquad
f^z = \kappa r H^r H^{(0.5)}_s;
\end{equation}
\end{itemize}

\item
the finite-difference analogue of the additional conservation law~(\ref{special_CL_1}) for the case~$\sigma=C\rho$
\begin{multline}
\left[
    \left(2 \check{t} - \frac{C s}{\kappa}\right) \frac{H}{\rho}
    - \frac{C}{\kappa} r H^r z_-
\right]_t
\\
\; + \left[
    \left(2 (t - \lambda\tau) - \frac{C s_-}{\kappa} \right)
    \left({\mathcal{E}}^{(\lambda)} - (r H^r w^{(0.5)})_-\right)
    - (r^2 H_-)^{(\lambda)}
\right]_s = 0.
\end{multline}
\end{itemize}

\subsubsection{Symmetry analysis of the constructed scheme}

Here we consider the symmetries of the constructed scheme in case of polytropic gas
with the state equation~(\ref{InternalEnergyRel}).
In this case, equation~(\ref{SamPopScheme1_energy_evol}) becomes
\begin{equation}
p_t = -\rho \, (\hat{p} + (\gamma - 1) p^{(\alpha)}) \, (r^{(0.5)}u^{(0.5)})_s + (\gamma - 1) \rho q,
\end{equation}
where $q$ is given by (\ref{SamPopScheme1_q}), and the
remaining equations of system~(\ref{SamPopScheme1}) do not change.

\medskip

The uniformness and orthogonality of the mesh
are preserved by a group transformation if
the following conditions for the corresponding generator
are satisfied~\cite{Dor_1,bk:Dorodnitsyn[2011]}
\begin{equation} \label{mesh_conds_uni}
  \dsp{D}\dsm{D}(\xi^s) = 0,
  \qquad
  \dtp{D}\dtm{D}(\xi^t) = 0,
\end{equation}
\begin{equation} \label{mesh_conds_ortho}
  \underset{\pm{s}}{D}(\xi^t) = -\underset{\pm{\tau}}{D}(\xi^s).
\end{equation}
The latter conditions are held for all the generators~(\ref{kern01}),
(\ref{kern01_ext1}), (\ref{kern02}), (\ref{kern02_ext1}) and~(\ref{kern02_ext2}).

\medskip

One can verify that the symmetries admitted by the schemes are the same as
for the corresponding systems of differential equations. Following Section~\ref{sec:syms},
we consider two cases separately:

\begin{enumerate}[label=\arabic*)]

\item
\textbf{Case $A \neq 0$}

System~(\ref{SamPopSchemeExt1}), (\ref{InternalEnergyRel}) admits the generators~(\ref{kern01}) for arbitrary~$\sigma$, and the generator~(\ref{kern01_ext1}) for~$\sigma=C\rho$.

\item
\textbf{Case $A = 0$}

System~(\ref{SamPopSchemeExt1}), (\ref{InternalEnergyRel}) admits the generators~(\ref{kern02}) for arbitrary~$\sigma$,
and the generators~(\ref{kern02_ext1}), (\ref{kern02_ext2}) for~$\sigma=C\rho$.

Notice that the generator~$\displaystyle X_3^0 =  f( s, r v ) {\ddtheta}$ is admitted by
the last equation of~(\ref{SamPopSchemeExt1}) in virtue of~(\ref{angularMomCL_triv}):
\begin{equation}
X_3^0\left(\theta_t - \frac{v^{(0.5)}}{r^{(0.5)}}\right)
= \frac{1}{\tau}(f(s, \hat{r}\hat{v}) - f(s, rv))\bigg|_{(\ref{angularMomCL_triv})} = 0.
\end{equation}

\end{enumerate}


\section{Schemes for the case of infinite conductivity}
\label{sec:sec4}

\label{sec:inf}

Limiting $\sigma \to \infty$ in (\ref{Lagrange_system}), one derives
\begin{subequations}    \label{Lagrange_system3}
\begin{gather}
\rho_t = -\rho^2 (r u) _{{s}},
  \label{Lagrange_system3_rho}
\\
 u_t   -  { v^2 \over  r }  = - r p_{{s}}
-   { 1  \over 2  r  }       \left(   r^2  ( H^{\theta} )^2     \right)  _s
-   { r  \over 2 }       \left(    ( H^z )^2     \right)  _s  ,
 \label{Lagrange_system3_u}
\\
 v _t   +   {  u v  \over  r }
=  { H^r  }     \left(   r   H^{\theta}    \right)  _s ,
\qquad
{\theta} _t  = {  v \over r }    ,
 \label{Lagrange_system3_v}
\\
 w_t
=  r { H^r }         H^{z}     _s ,
\qquad
z_t =  w ,
 \label{Lagrange_system3_w}
\\
\varepsilon_t = -p (r u)_s,
  \label{Lagrange_system3_p}
\\
  H^{\theta}  _t
=
r \rho (   ( v H^r  )  _s    -  H^{\theta} u_s )  ,
 \label{Lagrange_system3_H_theta}
\\
  H^z _t
=  \rho  (     ( r  w H^r  ) _s      -   H^z  (r u )_s   )
  \label{Lagrange_system3_H_z}
\end{gather}
\end{subequations}
together with~(\ref{nonlocal_r}).

Recall that in the case of infinite conductivity, the electric field vector vanishes~\cite{bk:KulikovskiiLubimov1965,bk:SamarskyPopov_book[1992]}, i.e.,
\begin{equation} \label{EisZero}
E^\theta \equiv 0,
\qquad
E^z \equiv 0.
\end{equation}


\medskip

For the polytropic gas with the state equation~(\ref{InternalEnergyRel}), taking into account~(\ref{Lagrange_system3_rho}), the energy evolution equation~(\ref{Lagrange_system3_p}) is reduced to
\begin{equation} \label{Lagrange_system3_S}
\left(\frac{p}{\rho^\gamma}\right)_t = S_t = 0.
\end{equation}
This means the preservation of the entropy~$S=S(s)$ along trajectories of motion.

One can also rewrite~(\ref{Lagrange_system3_S}) in evolutionary form as
\begin{equation}\label{Lagrange_system3_S_p_evol}
p_t = -\gamma  \rho p (r u)_{s}.
\end{equation}
Equation~(\ref{Lagrange_system3_S_p_evol}) is referred further when constructing
entropy-preserving finite difference schemes.

\subsection{Symmetries admitted by system~(\ref{Lagrange_system3}), (\ref{nonlocal_r})}

In the present section, the symmetries admitted by system~(\ref{Lagrange_system3}), (\ref{nonlocal_r})
are given according to~\cite{bk:DorKozMelKap_CylFlows_2022}.

\smallskip

In \textbf{case} $A \neq 0$, the system admits the following Lie algebra
\begin{multline} \label{EqnInfSymsAneq0}
X_1 = {\ddt},
\qquad
X_2 = {\dds},
\qquad
X_3 = t  {\ddt} + 2 s {\dds}
    - u {\ddu} - v {\ddv} - w {\ddw}
    + 2 \rho {\ddrho},
\\
X_4 =
- 2 s {\dds} +  r {\ddr}  +  z {\ddz}
    +  v {\ddv} +   u {\ddu} + w {\ddw}
    -  4 \rho {\ddrho}    -  2 p {\ddp}
 -  H^{\theta}   {\ddHtheta} -  H^z  {\ddHz} ,
\\
X_{5} = t {\ddz} + {\ddw}  ,
\qquad
X_6 =    f_1   \left(s, { p  \over \rho^{\gamma} }  \right)    {\ddtheta} ,
\qquad
X_{7} = f_2 \left(s, { p  \over \rho^{\gamma} }  \right)    {\ddz},
\end{multline}
where $ f_1 $ and $ f_2 $ are arbitrary functions of their arguments.

\bigskip

In \textbf{case} $A = 0$, the system admits the generators
\begin{multline} \label{EqnInfSymsAeq0}
X_1 = {\ddt},
\qquad
X_2 = {\dds},
\qquad
X_3 = t  {\ddt} + 2 s {\dds}
    - u {\ddu} - v {\ddv}
    + 2 \rho {\ddrho},
\\
X_4 =
- 2  s {\dds} +  r {\ddr}
    +  v {\ddv} +   u {\ddu}
    -  4 \rho {\ddrho}  -  2 p {\ddp}
-  H^{\theta}   {\ddHtheta} -  H^z  {\ddHz} ,
\\
X_5 =
  2  s {\dds}
    + 2  \rho {\ddrho}  +   2 p {\ddp}
  +  H^{\theta}   {\ddHtheta}  +   H^z  {\ddHz}  ,
\qquad
X_6 =    g_1    {\ddtheta} ,
\end{multline}
where
\begin{equation*}
g_1 =  g_1 \left(   s, r v,   { p \over \rho ^{ \gamma}  } ,   { H^{\theta}  \over r \rho}   ,  { H^{z}  \over \rho}  \right)
\end{equation*}
is an arbitrary function of its arguments.

For $\gamma=2$ there is an extension by the generator
\begin{equation} \label{X7forg2}
X_7 =
 \rho    g_{2}
 \left(   {\ddHz}    -    H^{z}    {\ddp}   \right),
\end{equation}
where
\begin{equation*}
g_{2} = g_2 \left(   s, r v,   { p \over \rho ^{ \gamma}  } ,
{ H^{\theta}  \over r \rho}   ,  { H^{z}  \over \rho}  \right)
\end{equation*}
is an arbitrary function.

\medskip

One notes that for $A=0$ part of equations (\ref{Lagrange_system3}) can be integrated
\begin{equation}
 S=S(s) ,
\qquad
 H^{\theta}= r \rho F(s),
\qquad
H^z = \rho G(s),
\qquad
v = R(s)/ r,
\end{equation}
where the functions $S(s)$, $F(s)$, $G(s)$ and $R(s)$ are arbitrary. As fixing the arbitrary elements leads to the extension of admitted Lie group, then in~\cite{bk:DorKozMelKap_CylFlows_2022} group classification of the~MHD equations
in~Lagrangian coordinates~(\ref{Lagrange_system3}) with respect to these functions was performed.


\subsection{Conservation laws possessed by system~(\ref{Lagrange_system3}), (\ref{nonlocal_r})}
\label{sec:add_CLs_inf}

Assuming~(\ref{EisZero}), the conservation laws
(\ref{CL_general_A_mass}), ..., (\ref{CL_general_A_energy})
are preserved for~$A \neq 0$, and the conservation laws
(\ref{CL_A0_mass}), ..., (\ref{CL_A0_energy})
are preserved for~$A = 0$.

In case $A=0$, there is also an infinite set of conservation laws of the form
\begin{equation} \label{InfSetOfCLs}
D_t \left\{
 \Phi  \left(  r v,     S,  { H^{\theta}   \over  r \rho   }  ,   {  H^z   \over  \rho }  ,  w, z - t w  \right)
\right\} = 0,
\end{equation}
where $\Phi$ is an arbitrary differentiable function of its arguments.

In addition, there are numerous conservation laws that arise for specific forms of the functions~$S$, $F$, $G$, and~$R$.
They were obtained in~\cite{bk:DorKozMelKap_CylFlows_2022} and are briefly listed below.

\bigskip

\textbf{In case} $A \neq 0$, there are only two additional conservation laws.
\begin{itemize}

\item
For $S(s)=S_0 = \text{const}$,

\begin{equation}     \label{conservation_general_A_s_physical}
D_t
\left(
  { u   H^r      +     v  H^{\theta}    +     w H^z    \over  r \rho  H^r }
\right)
+
D_s
\left(
 -
{ 1   \over 2 }    (    u  ^2 + v ^2 +  w^2  )
 +
  {  \gamma S \over  \gamma  -1  }    \rho  ^{ \gamma -1 }
\right)
= 0   .
\end{equation}

\item
For $ S  (s) = S_0  s^q $ with $  q = 1 - 2 \gamma$,
\begin{multline}      \label{CL_scaling_with _A}
D_t
\left\{
2 s
{ u   H^r      +     v  H^{\theta}    +     w H^z    \over  r \rho  H^r }
+
r u
+
z w
\right\}
\\
+
D_s
\left\{
2s
\left(
 -  { 1   \over 2 }    (    u  ^2 + v ^2 +  w^2  )
 +         { \gamma  S \over  \gamma  -1  }    \rho  ^{ \gamma -1 }
\right)
\right.
\\
\left.
+
r ^2
\left(
  {  S    }    \rho  ^{\gamma}
+
{    (  H ^{\theta} )^2    +    (  H^z ) ^2 \over 2 }
\right)
-
 r
{  H^r  }
z
  H ^z
\right\} = 0.
\end{multline}

\end{itemize}

\textbf{In case} $A = 0$,
the following additional conservation laws arise depending on the form of the functions $S$, $F$, $G$, and $R$.

\begin{itemize}

  \item
  For all constant values  $ S=S _0  $,       $ F=F _0  $,     $ G=G _0  $  and    $ R=R_0$,
\begin{equation}       \label{conservation_A_s_physical}
D_t
\left(
{    u   \over r \rho  }
\right)
+
D_s
\left(
{    1  \over 2 }     (  - u  ^2  + v ^2   )
+         {  \gamma  S    \over    \gamma  -1  }  \rho  ^{\gamma -1}
+          {  ( H^{\theta}  ) ^2   +   ( H^z ) ^2   \over    \rho }
\right)
= 0.
\end{equation}

\item
For the specified constants
$\gamma =  { 5 \over 4 } $, $ F_0 \neq 0 $,   $ G_0 = 0 $,  $ R_0  = 0 $ ($S_0$ is any),
\begin{multline}
D_t
\left\{
 4  t
\left(
   {  u ^{2}  \over 2 }
+  {S  \over  \gamma - 1 }     \rho  ^{\gamma - 1 }
+           {  ( H^{\theta}  ) ^2   \over  2   \rho }
 \right)
 -
 2   s
   { u \over r  \rho }
 -
 3    r
 u    \right\}
\\
+
D_s
\left\{
   (   4  t  r    u
-
 3   r^2 )
\left(
 S   \rho ^{\gamma}
+     {  ( H^{\theta}  ) ^2     \over  2   }   \right)
\right.
\\
\left.
  -
2   s
\left(
 -   { u ^{2}    \over 2 }
+   { \gamma  S \over  \gamma - 1 }     \rho ^{\gamma -1}
+       {  ( H^{\theta}  ) ^2   \over    \rho }
 \right)
\right\} = 0  .
\end{multline}

\item
For
$S = S_{0} s^{q_1}  ,
F = F_{0} s^{q_2}  ,
G = G_{0} s^{q_3}  ,
R = R_{0} s^{q_4},$
provided that $F_0 \neq 0$ and
\begin{equation*}
q_1 =   - 2  ( \gamma -2 ) q_2 - 4 \gamma + 5   ,
\qquad
q_3 = - { 3  \over 2  } ,
\qquad
q_4 = -1,
\end{equation*}
there is the conservation law
\begin{multline}     \label{CL_power_A_2_1}
D_t
\left\{
  4  (  q_2 +  1  )     t
\left(
   {  u ^{2}  +  v ^2  \over 2 }
+   {S  \over  \gamma - 1 }     \rho  ^{\gamma - 1 }
+          {  ( H^{\theta}  ) ^2   +   ( H^z ) ^2   \over  2   \rho }
 \right)
\right.
\\
\left.
 -
2  s
   { u \over r  \rho }
 -
  ( 2 q _2  + 3  )  r
   u
\right\}
\\
+
D_s
\left\{
  ( 4  (  q_2 +  1  )   t    r u
-
   ( 2 q _2  + 3  )  r ^2    )
\left(
   S   \rho ^{\gamma}
+     {  ( H^{\theta}  ) ^2   +   ( H^z ) ^2   \over  2   }
\right)
\right.
\\
\left.
 -
2  s
\left(
   { - u ^{2}  +  v ^2    \over 2 }
+   { \gamma  S \over  \gamma - 1 }     \rho ^{\gamma -1}
+        {  ( H^{\theta}  ) ^2   +   ( H^z ) ^2   \over    \rho }
 \right)
\right\}= 0   .
\end{multline}

\item
For
$S = S_{0} s^{q_1}  ,
F = F_{0} s^{q_2}  ,
G = G_{0} s^{q_3}  ,
R = R_{0} s^{q_4},$
provided that  $ F_0 = 0  $,    $ G_0 \neq 0 $, and
\begin{equation*}
q_3 = - { 3  \over 2  } ,
\qquad
q_4 = -1  .
\end{equation*}
there is the conservation law
\begin{multline}
D_t
\left\{
  2  (  q_1 + 2 \gamma  - 1  )    t
\left(
  {  u ^{2}  +  v ^2  \over 2 }
+   {S  \over  \gamma - 1 }     \rho  ^{\gamma - 1 }
+           {   ( H^z ) ^2   \over  2   \rho }
 \right)
\right.
\\
\left.
  +
  2 ( \gamma - 2 ) s
   { u \over r  \rho }
-
 ( q_1 + \gamma   + 1  )   r
   u
\right\}
\\
+
D_s
\left\{
  (    2   (  q_1 + 2 \gamma   - 1  )    t   r u
-
  (   q_1 +  \gamma + 1  )   r^2  )
\left(
  S   \rho ^{\gamma}
+    {  ( H^z ) ^2   \over  2   }   \right)
\right.
\\
  +
\left.
 2 ( \gamma - 2 ) s
\left(
   { - u ^{2}  +  v ^2    \over 2 }
+  { \gamma  S \over  \gamma - 1 }     \rho ^{\gamma -1}
+       {     ( H^z ) ^2   \over    \rho }
 \right)
\right\}= 0  .
\end{multline}

\item
For $S (s) =S_{0} e^{q_1 s}  ,
F (s) = F_{0} e^{q_2 s} ,
G(s) =G_{0} e^{q_3 s}  ,
R (s) =R_{0} e^{q_4 s}$,
provided that $ F_0 \neq 0 $ and
\begin{equation*}
q_1 = - {   2 ( \gamma - 2  )  q_2  } ,
\qquad
q_3 = q_4 =  0,
\end{equation*}
there is the conservation law
\begin{multline}       \label{CL_power_A_3_1}
D_t
\left\{
    2  q_2     t
\left(
   {  u ^{2}  +  v ^2  \over 2 }
+   {S  \over  \gamma - 1 }     \rho  ^{\gamma - 1 }
+          {  ( H^{\theta}  ) ^2   +   ( H^z ) ^2   \over  2   \rho }
 \right)
 -
  { u \over r  \rho }
-
  q _2  r
  u    \right\}
\\
+
D_s
\left\{
 q_2   (  2    t  r   u
-
  r^2 )
  \left(
 S   \rho ^{\gamma}
+   {  ( H^{\theta}  ) ^2   +   ( H^z ) ^2   \over  2   }   \right)
\right.
\\
\left.
 -
\left(
   { - u ^{2}  + v ^2    \over 2 }
+  { \gamma  S \over  \gamma - 1 }     \rho ^{\gamma -1}
+       {  ( H^{\theta}  ) ^2   +   ( H^z ) ^2   \over    \rho }
 \right)
\right\}= 0 .
\end{multline}

\item
For $S (s) =S_{0} e^{q_1 s}  ,
F (s) = F_{0} e^{q_2 s} ,
G(s) =G_{0} e^{q_3 s}  ,
R (s) =R_{0} e^{q_4 s}$,
provided that
$ F_0 = 0 $,   $ G_0  \neq  0 $, and
\begin{equation*}
q_3 = q_4 =  0,
\end{equation*}
there is the conservation law
\begin{multline}
D_t
\left\{
    2  q_1     t
\left(
  {  u ^{2}  +  v ^2  \over 2 }
+   {S  \over  \gamma - 1 }     \rho  ^{\gamma - 1 }
+           {     ( H^z ) ^2   \over  2   \rho }
 \right)
  +
2  (\gamma - 2 )
 { u \over r  \rho }
-
q _1   r      u
\right\}
\\
+ D_s
\left\{
q_1  (     2      t   r  u
-
  r   ^2  )
  \left(
  S   \rho ^{\gamma}
+      {  ( H^z ) ^2   \over  2   }   \right)
\right.
\\
\left.
  +
  2  (\gamma - 2 )
\left(
   { - u ^{2}  +  v ^2    \over 2 }
+   { \gamma  S \over  \gamma - 1 }     \rho ^{\gamma -1}
+        {     ( H^z ) ^2   \over    \rho }
 \right)
\right\} = 0 .
\end{multline}

\item
For arbitrary $S_1$, $F$, $R$, where
\begin{equation}     \label{define_tilde_S}
S_1 = {   S \over \gamma -1}    +  { G ^2  \over 2 },
\end{equation}
provided $ F(s) \equiv 0 $, $  G(s) \ {\not \equiv } \ 0  $, and $\gamma=2$,
one gets
\begin{multline}     \label{CL_gamma_1}
D_t
\left\{
   2 t
\left(
  {  u ^{2}  +  v ^2  \over 2 }
+   {S  \over  \gamma - 1 }     \rho  ^{\gamma - 1 }
+          {    ( H^z ) ^2   \over  2   \rho }
 \right)
  -   r
    u
\right\}
\\
+
D_s
\left\{
  (  2 t  r    u
-
 r ^2 )
  \left(
  S   \rho ^{\gamma}
+     {    ( H^z ) ^2   \over  2   }   \right)
\right\}
= 0
\end{multline}
and
\begin{multline}      \label{CL_gamma_2}
D_t
\left\{
   t ^2
\left(
   {  u ^{2}  +  v ^2  \over 2 }
+  {S  \over  \gamma - 1 }     \rho  ^{\gamma - 1 }
+           {     ( H^z ) ^2   \over  2   \rho }
 \right)
-
 t  r    u
+
 { r  ^2 \over 2 }
\right\}
\\
+
D_s
\left\{
 (   t ^2    r      u
  -
 t    r^2   )
   \left(
 S   \rho ^{\gamma}
+     {     ( H^z ) ^2   \over  2   }   \right)
\right\}
= 0  .
\end{multline}

\item
For $S_1 (s) = {S}_{0} s^{q_1},
F (s) = F_{0} s^{q_2}  ,
R (s) =R_{0} s^{q_3}$,
where $S_1$ is given by~(\ref{define_tilde_S}),
provided that
$ F(s) \equiv 0 $, $  G(s) \ {\not \equiv } \ 0  $, $\gamma=2$, and
\begin{equation*}
q_1  = -3 ,
\qquad
q_3 = -1,
\end{equation*}
there is the conservation law
\begin{multline}
D_t
\left\{
   t
\left(
   {  u ^{2}  +  v ^2  \over 2 }
+   {S  \over  \gamma - 1 }     \rho  ^{\gamma - 1 }
+           {     ( H^z ) ^2   \over  2   \rho }
 \right)
  +
  s
   { u \over r  \rho }
\right\}
\\
+
 D_s
\left\{
    t
  r   u    \left(
    S   \rho ^{\gamma}
+       {     ( H^z ) ^2   \over  2   }   \right)
  +  s
\left(
   {  - u ^{2}  +  v ^2    \over 2 }
+  { \gamma  S \over  \gamma - 1 }     \rho ^{\gamma -1}
+        {     ( H^z ) ^2   \over    \rho }
 \right)
\right\} = 0   .
\end{multline}

\end{itemize}

\subsection{Numerical schemes for system~(\ref{Lagrange_system3}), (\ref{nonlocal_r})}

Modifying scheme~(\ref{SamPopSchemeExt1}) in a similar way as it was done for system~(\ref{Lagrange_system3}), one derives the following scheme.
\begin{subequations} \label{Inf_SamPopSchemeExt1}
\begin{equation} \label{Inf_SamPopSchemeExt1_mass}
\left(\frac{1}{\rho}\right)_t = (r^{(0.5)} u^{(0.5)})_s,
\end{equation}
\begin{equation} \label{Inf_SamPopSchemeExt1_u}
u_t - \frac{\hat{v} v^{(0.5)}}{r} + r^{(0.5)} p^{(\alpha)}_{\bar{s}} - f^r = 0,
\end{equation}
\begin{equation} \label{Inf_SamPopSchemeExt1_v}
v_t + \frac{\hat{v} u^{(0.5)}}{r}
    - \kappa H^r (h^\theta)^{(0.5)}_s = 0,
\end{equation}
\begin{equation}
w_t - \kappa \left( r H^r H^{(0.5)} \right)_s = 0,
\end{equation}
\begin{equation} \label{Inf_SamPopScheme1_CL_Hz_ext}
\left(\frac{H}{\rho}\right)_t
    - (r H^r w^{(0.5)})_{\bar{s}}
= 0,
\end{equation}
\begin{equation} \label{SamPopScheme1_CL_Hy_ext}
\left(\frac{h^\theta}{\rho (r^*_+)^2}\right)_t
        + (H^r v^{(0.5)})_{\bar{s}}
= 0,
\end{equation}
\begin{equation} \label{Inf_SamPopScheme1_energy_evol}
\varepsilon_t = -p^{(\alpha)} (r^{(0.5)} u^{(0.5)})_s,
\end{equation}
\begin{equation}
r_t = u^{(0.5)},
\qquad
r_s = \frac{2}{(r + r_+) \rho},
\end{equation}
\begin{equation}
z_t = w^{(0.5)},
\qquad
\theta_t = \frac{v^{(0.5)}}{r^{(0.5)}},
\end{equation}
\end{subequations}
where $f^r$, $h^\theta$, and $H$ are given by (\ref{SamPopScheme1_fr}) and~(\ref{SamPopSchemeAbbr}).

\medskip

Since scheme~(\ref{Inf_SamPopSchemeExt1}) is inherited from scheme~(\ref{SamPopScheme1}), all the conservation laws given for scheme~(\ref{SamPopScheme1}) in the case of arbitrary~$\sigma$ are also valid for~(\ref{Inf_SamPopSchemeExt1}), assuming~(\ref{EisZero}).

\medskip

In the case of a polytropic gas, one should also modify equation~(\ref{Inf_SamPopScheme1_energy_evol})
so that it describes the conservation of entropy along trajectories of motion. This problem was recently studied by the authors in~\cite{bk:DorKapMDPI2022} for plane one-dimensional~MHD flows.
As equations~(\ref{Inf_SamPopSchemeExt1_mass}) and~(\ref{Inf_SamPopScheme1_energy_evol}) are similar to the finite-difference evolution equations for density end energy considered in~\cite{bk:DorKapMDPI2022}, the previously obtained results are easily carried over to the case of scheme~(\ref{Inf_SamPopSchemeExt1}).
Here we briefly recall the main results of~\cite{bk:DorKapMDPI2022} related to the entropy preservation.

\smallskip

The problem is to bring the nondivergent equation~(\ref{Inf_SamPopScheme1_energy_evol}) to a divergent finite-difference approximation of~(\ref{Lagrange_system3_S}), having the meaning of the entropy preservation along trajectories of motion.
We show that in the finite-difference case this can be done by choosing an appropriate approximation for the equation of state~(\ref{InternalEnergyRel}). First, we consider the particular case~$\gamma=2$.
We seek for an approximation of the form $\varepsilon = p^{(\alpha)} \eta$, where $\eta \sim 1/((\gamma-1)\rho)$.
Substituting into~(\ref{Inf_SamPopScheme1_energy_evol}) and taking into account~(\ref{Inf_SamPopSchemeExt1_mass}), we get
\begin{equation} \label{G2approx1}
\displaystyle
\frac{p^{(\alpha)}}{\check{p}^{(\alpha)}}
= \frac{\rho - \check{\rho} + \rho\check{\rho} \check{\eta}} {\rho\check{\rho} \eta}.
\end{equation}
The right hand side can be reduced to a divergent expression in case $\eta=1/{\hat{\rho}}$.
Namely, equation~(\ref{G2approx1}) becomes
\begin{equation}
\frac{p^{(\alpha)}}{\check{p}^{(\alpha)}} = \frac{\rho\hat{\rho}}{\check{\rho}\rho},
\end{equation}
or
\begin{equation}
\left(\ln \check{p}^{(\alpha)} \right)_t =
    \left(\ln \rho \check{\rho} \right)_t.
\end{equation}
The latter equation is equivalent to the following finite-difference derivative
\begin{equation}\label{schemeEntropyCLinf2}
\twopointS_t = \left(\frac{\check{p}^{(\alpha)}}{\rho\check{\rho}}\right)_t = 0,
\end{equation}
where $\twopointS$ is a two-point representation of the entropy.

In a similar way conservation of entropy for~$\gamma=3$ is derived.
Choosing the approximation
\begin{equation}
\varepsilon = \frac{\rho p^{(\alpha)}}{\hat{\rho}(\hat{\rho} + \rho)}
\end{equation}
and substituting it into~(\ref{Inf_SamPopScheme1_energy_evol}), one derives
\begin{equation}
\twopointS_t = \left(\frac{2 \check{p}^{(\alpha)}}{\rho\check{\rho}(\rho + \check{\rho})}\right)_t = 0.
\end{equation}
Continuing the procedure for $\gamma=4, 5, \dots$, we establish the
following formulas for any natural~$\gamma \geqslant 2$.
\begin{equation} \label{schemeEntropyCLinfN}
\displaystyle
\varepsilon = \frac{p^{(\alpha)}}{\sum_{k=0}^{\gamma-2} \hat{\rho}^{\gamma-k-1} {\rho}^{k - \gamma + 2}},
\qquad
\twopointS_t = \left(\frac{(\gamma-1)\check{p}^{(\alpha)}}{\sum_{k=0}^{\gamma-2} \rho^{\gamma-k-1} \check{\rho}^{k+1} }\right)_t = 0.
\end{equation}
Similar formulas can be obtained for rational values~$\displaystyle \gamma = \frac{n}{m}$, where $n, m \in \mathbb{N}$ and $n > m$:
\begin{equation}
\displaystyle
\varepsilon = \frac{p^{(\alpha)} \rho^{\gamma-2} \, \mathcal{B}_{\rho}(m-1, m)}{\hat{\rho} \, \mathcal{B}_{\rho}(n-m-1, m)},
\qquad
\twopointS_t = \left(\frac{(\gamma-1)\check{p}^{(\alpha)} \check{\mathcal{B}}_{\rho}(m-1, m)}
{\rho \check{\rho} \, \check{\mathcal{B}}_{\rho}(n-m-1, m)}\right)_t = 0,
\end{equation}
where
\begin{equation}
\displaystyle
\mathcal{B}_{\rho}(n, m) = \sum_{k=0}^n \hat{\rho}^{\frac{n-k}{m}} \rho^{\frac{k}{m}}.
\end{equation}
Here we also provide the formulas for the specific cases~$\gamma = 5/3$ (one-atomic ideal gas) and $\gamma=7/5=1.4$ (diatomic gas), which often occur in applications.
\begin{equation}
\gamma=\frac{5}{3}:
\quad
\varepsilon = {p}^{(\alpha)}\, \frac{\hat{\rho}^{2/3} + (\rho\hat{\rho})^{1/3} + \rho^{2/3}}{\rho^{1/3}\hat{\rho}(\hat{\rho}^{1/3} + \rho^{1/3})},
\qquad
\twopointS_t =
\left(
\frac{2}{3}\,
\check{p}^{(\alpha)}\, \frac{\check{\rho}^{2/3} + (\rho\check{\rho})^{1/3} + \rho^{2/3}}{\rho\check{\rho}(\check{\rho}^{1/3} + \rho^{1/3})}
\right)_t = 0;
\end{equation}
\begin{multline}
\gamma=\frac{7}{5}:
\quad
\varepsilon = {p}^{(\alpha)}\, \frac{
        \hat{\rho}^{4/5}
        + \hat{\rho}^{3/5} \rho^{1/5}
        + (\hat{\rho} \rho)^{2/5}
        + \hat{\rho}^{1/5} \rho^{3/5}
        + \rho^{4/5}
    }
    {
        \rho^{3/5}\hat{\rho}\, (\hat{\rho}^{1/5} + \rho^{1/5})
    },
\\
\twopointS_t =
\left(
\frac{2}{5}\,
\check{p}^{(\alpha)}\,
    \frac{
        \check{\rho}^{4/5}
        + \check{\rho}^{3/5} \rho^{1/5}
        + (\check{\rho} \rho)^{2/5}
        + \check{\rho}^{1/5} \rho^{3/5}
        + \rho^{4/5}
    }
    {
        \rho \check{\rho}\, (\check{\rho}^{1/5} + \rho^{1/5})
    }
\right)_t = 0.
\end{multline}

\bigskip

In case~$H^r=0$ ($A=0$), there are also conservation laws that arise in the finite-difference case similarly to~(\ref{InfSetOfCLs}). For $H^r=0$ the quantities
\[
r v,
\quad
\frac{H}{\rho},
\quad
\frac{h^\theta}{\rho (r^*_+)^2},
\quad
w,
\quad
z - t w^{(0.5)}
\]
are preserved along the trajectories, as it can be seen from system~(\ref{Inf_SamPopSchemeExt1}).
Thus, there is an infinite set of conservation laws of the form
\begin{equation}
\left\{
 \widetilde{\Phi}\left(r v, \twopointS, \frac{H}{\rho} , \frac{h^\theta}{\rho (r^*_+)^2}, w, z - t w^{(0.5)}\right)
\right\}_t = 0,
\end{equation}
where $\widetilde{\Phi}$ is an arbitrary function of its arguments.

\medskip

Studying the invariance of scheme~(\ref{Inf_SamPopSchemeExt1}), one verifies that in case $A\neq 0$
the scheme admits the generators~$X_1$, ..., $X_5$ of~Lie algebra~(\ref{EqnInfSymsAneq0}).
However, the scheme is not invariant with respect to the generators~$X_6$ and~$X_7$,
and one should consider the generators
\begin{equation}
\widetilde{X}_6 = f_1   \left(s, \twopointS\right) {\ddtheta},
\qquad
\widetilde{X}_7 = f_2 \left(s, \twopointS \right) {\ddz},
\end{equation}
instead, where $\twopointS$ is the chosen two-point approximation for the entropy.

\smallskip

In case $A=0$, scheme~(\ref{Inf_SamPopSchemeExt1}) is invariant with respect to the generators~$X_1$, ..., $X_5$ of~Lie algebra~(\ref{EqnInfSymsAeq0}).
Similar to the case~$A \neq 0$ one should consider the generator
\begin{equation}
\widetilde{X}_6 = g_1\left(s, r v, \twopointS, \frac{H}{\rho} , \frac{h^\theta}{\rho (r^*_+)^2}\right){\ddtheta},
\end{equation}
instead of $X_6$.
Finally, the generator $X_7$ (for $\gamma=2$ only) is not admitted by equations~(\ref{Inf_SamPopSchemeExt1_u}) and~(\ref{Inf_SamPopScheme1_energy_evol}).
This could be expected, since the construction of the scheme was carried out on the
basis of conservation laws, while there is no conservation law associated
with the generator~$X_7$~\cite{bk:DorKozMelKap_CylFlows_2022}.

\bigskip

Now we discuss the additional conservation laws that arise for various specific forms of the functions~$F$, $G$, $S$, and~$R$.
It turns out to be impossible to construct a scheme based on~(\ref{Inf_SamPopSchemeExt1}) that possesses difference analogues of the conservation laws listed in Section~\ref{sec:add_CLs_inf} in terms of rational difference expressions.
As an illustration, consider the conservation law~(\ref{conservation_A_s_physical})
\begin{multline} \label{conservation_A_s_physical_ii}
D_t
\left(
{    u   \over r \rho  }
\right)
+
D_s
\left(
{    1  \over 2 }     (  - u  ^2  + v ^2   )
+         {  \gamma  S_0    \over    \gamma  -1  }  \rho  ^{\gamma -1}
+          {  ( H^{\theta}  ) ^2   +   ( H^z ) ^2   \over    \rho }
\right) =
\\
= D_t
\left({u   \over r \rho}\right)
+ D_s\left(
    \frac{R_0^2}{2r^2} - \frac{u^2}{2}
    + F_0^2 r^2 \rho
    + \rho G_0^2
    + \frac{\gamma S_0}{\gamma-1} \rho^{\gamma-1}
\right) = 0,
\end{multline}
which occurs in case $S=S _0$, $F=F _0$, $G=G _0$, and $R=R_0$,
where $S_0, F_0, G_0$ and $R_0$ are constant.

\begin{remark}
As far as the authors know, the conservation law~(\ref{conservation_A_s_physical_ii}) has no a definite name.
We also present it here in the following integral form, which may be more convenient for further analysis.
\begin{equation}
\displaystyle
\frac{\partial}{\partial{t}} \int_{s_0}^s \frac{u}{r \rho} \, ds
+ \left[
    \frac{R_0^2}{2r^2} - \frac{u^2}{2}
    + F_0^2 r^2 \rho
    + \rho G_0^2
    + \frac{\gamma S_0}{\gamma-1} \rho^{\gamma-1}
\right]_{s_0}^s = 0.
\end{equation}
In Eulerian coordinates the latter is
\begin{equation}
\displaystyle
\frac{\partial}{\partial{t}} \int_{r_0}^r u\, ds
+ \left[
    \frac{u^2}{2}
    + \frac{R_0^2}{2r^2}
    + F_0^2 r^2 \rho
    + \rho G_0^2
    + \frac{\gamma S_0}{\gamma-1} \rho^{\gamma-1}
\right]_{r_0}^r = 0.
\end{equation}
Recall that the Eulerian coordinate $r$ is defined in the mass Lagrangian
coordinates $(t,s)$ by~(\ref{nonlocal_r}).
\end{remark}

One rewrites the conservation law~(\ref{conservation_A_s_physical_ii}) in terms of equations~(\ref{Lagrange_system3})
and conservation law multipliers as follows
\begin{multline} \label{conservation_A_s_physical_expanded}
\frac{u}{r\rho^2} \left( \rho_t + \rho^2 (r u)_s \right)
    + \frac{u}{r^2 \rho} \left( r_t - u \right)
    + \left( \frac{R_0^2}{r^3} - \frac{u^2}{r} \right) \left(r_s - \frac{1}{r\rho}\right)
    \\
    - \frac{1}{r \rho} \left(
        u_t - \frac{R_0^2}{r^3}
            + \gamma S_0 r \rho^{\gamma-1} \rho_s
            + F_0^2 (r \rho_s + 2 \rho r_s) r^2 \rho
            + G_0^2 r \rho \rho_s
    \right) = 0.
\end{multline}
In the finite-difference case, the construction of such a conservation law for scheme~(\ref{Inf_SamPopSchemeExt1}) in rational expressions is not possible. This is explained by a large number of relations included in the finite-difference approximation of the conservation law~(\ref{conservation_A_s_physical_expanded}), which must be preserved in the difference case so that~(\ref{conservation_A_s_physical_expanded}) can be represented as a divergent expression. In this case, there is no freedom in choosing the approximation of the terms of the conservation law~(\ref{conservation_A_s_physical_expanded}), as we have already stated the form of the equations of the scheme.

\bigskip

An alternative approach is to construct new schemes based on reasonable approximations for the conservation law~(\ref{conservation_A_s_physical_expanded}). An example of such a scheme is the scheme
\begin{subequations} \label{explScheme}
\begin{equation}
\rho_t + \rho \hat{\rho} \,(\hat{r} u)_s = 0,
\end{equation}
\begin{equation}
(u_*^+)_t - \frac{R_0^2}{\hat{r} r^2} + \hat{r}\hat{\rho}\left(
    F_0^2 (r^2 \rho)_s
    + G_0^2 \rho_s
    + \frac{\gamma S_0}{\gamma-1} (\rho^{\gamma-1})_s
\right) = 0,
\end{equation}
\begin{equation}
v_t + \frac{v u_+}{\hat{r}} = 0,
\end{equation}
\begin{equation}
r_t = u_+,
\qquad
\hat{r}_s = \frac{1}{r \rho},
\end{equation}
\begin{equation}
\frac{H^\theta}{\rho} = F_0,
\qquad
\frac{H^z}{\rho} = G_0,
\end{equation}
\begin{equation}
r v = R_0,
\qquad
\frac{p}{\rho^\gamma} = S_0,
\end{equation}
\end{subequations}
which possesses the conservation law
\begin{multline} \label{expl_addit_cl_1}
\frac{u_*^+}{\hat{r} \rho \hat{\rho}} \left(
    \rho_t + \rho \hat{\rho} (\hat{r} u)_s
\right)
    + \frac{u_*^+}{\rho r \hat{r}} \left(
        r_t - u_+
    \right)
    + \left(
        R_0^2 \psi - \frac{u^+ u_*^+}{\hat{r}}
    \right) \left(\hat{r}_s - \frac{1}{r\rho} \right)
    \\
    - \frac{1}{\hat{r}\hat{\rho}} \left(
        (u_*^+)_t - \frac{R_0^2}{\hat{r} r^2} + \hat{r}\hat{\rho}\left(
            F_0^2 (r^2 \rho)_s
            + G_0^2 \rho_s
            + \frac{\gamma S_0}{\gamma-1} (\rho^{\gamma-1})_s
        \right)
    \right)
\\
= \left(
    \frac{u_*^+}{r\rho}
\right)_t + \left(
    \frac{R_0^2}{2 r^2}
    - \frac{u^2}{2}
    + F_0^2 r^2 \rho
    + \rho G_0^2
    + \frac{\gamma S_0}{\gamma-1} \rho^{\gamma-1}
\right)_s = 0,
\end{multline}
where
\begin{equation}
\psi = \frac{\rho}{2\hat{\rho} r \hat{r}^2 r_+^2} \frac{2 r_+^2 - \hat{\rho} \hat{r}^2 (r^2)_s}{1 - \rho r \hat{r}_s}
    = \frac{1}{r^3} + O(\tau + h).
\end{equation}

Notice that for any natural $\gamma>1$ one can write the evolution equation for the pressure~$p$ as
\begin{equation}
p_t = -\rho \hat{p} \left(\hat{r} u\right)_s \sum_{k=0}^{\gamma-1} \left(\frac{\rho}{\hat{\rho}}\right)^k,
\end{equation}
which is the finite-difference analogue of equation~(\ref{Lagrange_system3_S_p_evol}).
In case $\gamma$ is rational, the calculations become much more complicated and we consider only two particular cases:
\begin{align}
\displaystyle
& \gamma = \frac{5}{3}: \quad
p_t = -\rho \hat{p} \left(\hat{r} u\right)_s
    \left(1 + \frac{\rho}{\hat{\rho}} -
    \frac{\rho^{5/3}}{\hat{\rho}\left(\hat{\rho}^{2/3} + (\hat{\rho}\rho)^{1/3} + \rho^{2/3}\right)}
    \right);
\\
\displaystyle
& \gamma = \frac{7}{5}: \quad
p_t = -\rho \hat{p} \left(\hat{r} u\right)_s
    \left(1 + \frac{\rho}{\hat{\rho}}
    - \frac{\rho^{7/5} \, (\hat{\rho}^2 + \rho\hat{\rho} + \rho^2)}
        {
            \hat{\rho} \left(
                \hat{\rho}^{12/5}
                + \hat{\rho}^{9/5} \rho^{3/5}
                + (\hat{\rho} \rho)^{6/5}
                + \hat{\rho}^{3/5} \rho^{9/5}
                + \rho^{12/5}
            \right)
        }
    \right).
\end{align}

\medskip

According to its construction procedure, scheme~(\ref{explScheme}) possesses the following conservation laws
\begin{itemize}
\item

mass
\begin{equation}
\left(\frac{1}{\rho}\right)_t - (\hat{r} u)_s = 0;
\end{equation}

\item

angular momentum in $ (r, \theta )$-plane
\begin{equation}
(r v)_t = 0;
\end{equation}

\item

magnetic  fluxes
\begin{equation}
\left(\frac{H^\theta}{\rho}\right)_t = 0,
\qquad
\left(\frac{H^z}{\rho}\right)_t = 0;
\end{equation}

\item
entropy
\begin{equation}
\left(\frac{p}{\rho^\gamma}\right)_t = S_t = 0;
\end{equation}

\item
additional conservation law~(\ref{expl_addit_cl_1}).
\end{itemize}

Although the scheme does not possess the total energy conservation law,
it preserves the entropy~$S$ along trajectories of motion. A similar situation
was observed in~\cite{bk:DorKapMDPI2022} for plane magnetic flows.
Notice that, in contrast to the two-point representation of entropy~$\twopointS$
for scheme~(\ref{Inf_SamPopSchemeExt1}), here the entropy~$S$ is given at one point.

\medskip

Recall that the scheme has been constructed for the case when the
radial component of the magnetic field is absent~($A=0$).
Similarly, conservative schemes can be constructed for the case $A \neq 0$.
One of the possible schemes of this kind is given in~\ref{sec:apdxA}.

\section{Conclusion}

In the present paper, various finite-difference schemes for the one-dimensional~MHD equations in cylindrical geometry have been constructed.
All these schemes are based on the classical completely conservative~Samarskiy--Popov schemes. The schemes proposed in the paper generalize the Samarskiy--Popov schemes to the case when the magnetic field vector has a nonzero radial component, as well as to the case of frozen-in magnetic field, when the conductivity is infinite.

In the case of finite conductivity, it is shown that both the original~Samarskiy--Popov scheme and its extended version have a complete set of difference analogues of the conservation laws of the original differential models, including additional conservation laws that arise for a special form of conductivity. Extended schemes are still completely conservative ones, i.e., they possess finite-difference analogues of the conservation law of total energy,
together with the balance of internal and gas-dynamic energy.

In the case of a frozen-in magnetic field, mass, magnetic flux, momentum, angular momentum, the center-of-mass law, and total energy are preserved as in case conductivity is finite. It is also possible to preserve two-point representations of the entropy along the trajectories of motion.

As the analysis carried out in~\cite{bk:DorKozMelKap_CylFlows_2022} shows, the~MHD equations in the case of infinite conductivity possess numerous additional conservation laws that arise for special forms of entropy, magnetic flux, and other functions. The authors show that schemes of the~Samarskiy--Popov type have no difference analogues of these additional conservation laws.
Nevertheless, it is possible to construct specific finite-difference schemes that possess such conservation laws and also preserve entropy. An example of such a scheme and its conservation laws are given.

\section*{Acknowledgements}

The research was supported by Russian Science Foundation Grant No
18-11-00238 `Hydro\-dyna\-mics-type equations: symmetries, conservation
laws, invariant difference schemes'. E.I.K. acknowledges Suranaree
University of Technology~(SUT) and Thailand Science Research and
Innovation~(TSRI) for Full-time Doctoral Researcher Fellowship.


\appendix

\section{Conservative scheme for system~(\ref{Lagrange_system3}), (\ref{nonlocal_r}) in case $A \neq 0$}
\label{sec:apdxA}

In case the radial component of the magnetic field is present (i.e., $A \neq 0$),
the construction of conservative schemes for system (\ref{Lagrange_system3}), (\ref{nonlocal_r}) can be carried out according to the similar procedure as scheme~(\ref{explScheme}). Namely, possible approximations for the additional conservation law are considered, from the form of which a suitable family of schemes can be obtained. Further, the form of this family is refined in such a way that it possesses as many conservation laws as possible.

As in case $A\neq 0$ the calculations required to construct such a family of schemes become much more complicated compared to scheme~(\ref{explScheme}), the present section gives only the main results. Details can be found in~\cite{bk:KaptsovMHDCylSchemes_Conf2022}.

The resulting family of schemes has the following form
\begin{subequations} \label{A__Lagrange_system3_scheme}
\begin{equation}
\rho_t + \rho \hat{\rho} (\hat{r} u)_s = 0,
\end{equation}
\begin{equation}
(u_*^+)_t - \frac{\hat{r}\hat{\rho}}{r \rho} \frac{v_*^2}{r_{(1)}} + \hat{r} \hat{H}^\theta \Xi + \hat{r} \hat{H}^z (H^z_{(1)})_s
+ \frac{\gamma S_0}{\gamma-1} \hat{r}\hat{\rho} (\rho^{\gamma-1})_s
= 0,
\end{equation}
\begin{equation}
v^*_t + \frac{\hat{\rho}}{\rho} \frac{H^\theta}{\hat{H}^\theta} \frac{u v_*}{r_{(2)}} - A \Xi = 0,
\end{equation}
\begin{equation}
w^*_t - A (H^z_{(1)})_s = 0,
\end{equation}
\begin{equation}
r_t = u_+,
\qquad
\hat{r}_s = \frac{1}{r \rho},
\qquad
z_t = w_*,
\end{equation}
\begin{equation}
H^\theta_t + \hat{r}_+ \hat{\rho} \left(
    H^\theta u_s + \frac{A v^* \hat{r}_s}{\hat{r}_+ r_{(1)}}
    - \frac{A v_{\bar{s}}}{\hat{r}_+}
\right) = 0,
\end{equation}
\begin{equation}
H^z_t +\hat{\rho} \left(
    H^z (\hat{r} u)_s - A w_{\bar{s}}
\right) = 0,
\end{equation}
\begin{equation} \label{A__entropyEq}
\frac{p}{\rho^\gamma} = S_0,
\end{equation}
\end{subequations}
where $r_{(1)}$ and $r_{(2)}$ are some approximations for $r$,
$H^z_{(1)}$ is an approximation for $H^z$,
and $\Xi$ approximates the term~$\displaystyle \frac{1}{r}\,(r H^\theta)_s$.
These approximations are specified below.

The scheme preserves the entropy along trajectories of motion and possesses the following finite-difference analogues of the conservation laws (\ref{CL_general_A_mass}), ..., (\ref{CL_general_A_flux_z}), and~(\ref{conservation_general_A_s_physical}).
\begin{itemize}
    \item
    mass
    \begin{equation}
    \left(\frac{1}{\rho}\right)_t - (\hat{r} u)_s = 0;
    \end{equation}

    \item
    magnetic flux along $\theta$-axis
    \begin{equation}
    \left(\frac{H^\theta}{\rho r}\right)_t - \left(\frac{A v_-}{\hat{r}}\right)_s = 0,
    \end{equation}
    provided
    \[
        r_{(1)} = \frac{A v_*}{A v - h \hat{r}_+ H^\theta u_s} \, \hat{r}_+
            = r + O(h + \tau);
    \]

    \item
    magnetic flux along $z$-axis
    \begin{equation}
    \left(\frac{H^z}{\rho}\right)_t - (A w_-)_s = 0;
    \end{equation}

    \item
    momentum along $z$-axis
    \begin{equation}
    w^*_t - (A H^z_{(1)})_s = 0;
    \end{equation}

    \item
    motion of the center of mass along $z$-axis
    \begin{equation}
    (t w_* - z)_t - (A \hat{t} H^z_{(1)})_s = 0;
    \end{equation}

    \item
    angular momentum in $(r, \theta)$-plane
    \begin{equation}
    (v_* r_-)_t - (A r H^\theta)_s = 0,
    \end{equation}
    provided
    \[
    r_{(2)} = \frac{\hat{\rho}}{\rho} \frac{H^\theta}{\hat{H}^\theta} \, \hat{r}_-,
    \qquad
    \Xi = \frac{1}{\hat{r}_-} (r H^\theta)_s;
    \]

    \item
    entropy along trajectories of motion
    \begin{equation}
    \left(\frac{p}{\rho^\gamma}\right)_t = 0;
    \end{equation}

    \item
    difference analogue of the conservation law~(\ref{conservation_general_A_s_physical})
    \begin{equation}
    \left(
       \frac{u_*^+}{r \rho}
       + \frac{v_* H^\theta + w_* H^z}{A \rho}
    \right)_t
    - \left(
        \frac{u^2 + v^2_- + w^2_-}{2}
        - \frac{\gamma S_0}{\gamma-1} \rho^{\gamma-1}
    \right)_s = 0.
    \end{equation}
\end{itemize}
Notice that the conservation laws do not impose any restrictions on the choice of approximation~$H^z_{(1)}$.


\end{document}